\documentclass[pre,twocolumn,amsmath,amssymb,nofootinbib,floatfix,superscriptaddress]{revtex4}

\usepackage{graphicx,bm,float,physics}
\usepackage{dsfont}
\usepackage{xcolor}
\usepackage[normalem]{ulem}
\usepackage{placeins}

\begin{document}

\title{Stability of universal properties against perturbations of \\ the Markov Chain Monte Carlo algorithm}

\author{Matteo Bacci} 
\affiliation{Dipartimento di Fisica dell'Universit\`a di Pisa and 
INFN Sezione di Pisa, Largo Pontecorvo 3, I-56127 Pisa, Italy}

\author{Claudio Bonati} 
\affiliation{Dipartimento di Fisica dell'Universit\`a di Pisa and 
INFN Sezione di Pisa, Largo Pontecorvo 3, I-56127 Pisa, Italy}

\date{\today}

\begin{abstract}
We numerically investigate the stability of universal properties at continuous
phase transitions against perturbations of the Markov Chain Monte Carlo
algorithm used to simulate the system.  We consider the three dimensional XY
model as test bed, and both local (single site Metropolis) and global (single
cluster) updates, introducing deterministic truncation-like perturbations and
stochastic perturbations in the acceptance probabilities. In (almost) all the
cases we find a remarkable stability of the universal properties, even against
large perturbations of the Markov Chain Monte Carlo algorithm, with critical
exponents and scaling curves consistent with those of the standard XY model
within statistical uncertainties. Only for the single cluster update with very
large truncation error does something different happen, but large scaling
corrections prevent us from precisely assessing the critical properties of the
transition, and, in particular, to understand whether the critical behavior
observed corresponds to a known universality class.
\end{abstract}

\maketitle

\section{Introduction}
\label{intro}

Since only few statistical models can be solved analytically, and mainly in two
dimensions (see, e.~g., \cite{Baxter-book, IZ-book, McCoy-book}), in
statistical mechanics we often have to resort to numerical simulations.  The
Markov Chain Monte Carlo (MCMC) method is one of the most commonly used approaches
to numerically investigate the equilibrium properties of statistical systems
\cite{NB-book, BH-book, Berg-book, LB-book, Krauth-book, DELL-book}, since it
provides stochastically exact estimates of average values whenever its
application is not hindered by the presence of a sign problem \cite{LGSWSS-90,
Troyer:2004ge, Philipsen:2010gj, Aarts:2015tyj}.

Two different groups of applications of the MCMC method in statistical
physics can be naturally identified. In some cases we know exactly (or at
least with a high degree of accuracy) the microscopic details of the system we
are interested in, and our aim is to estimate some macroscopic properties of
direct physical interest, like, e.~g., the value of the specific heat for a
given temperature or the details of the phase diagram.  In other cases the
properties we want to study are not peculiar of a single physical model, but
characterize a whole class of systems, the typical example being that of
universal properties at continuous phase transitions \cite{WK-74, Fisher-75,
Ma-75, PV-02, ZJ-book}.  Since the critical behavior emerging close to a
continuous phase transition nonperturbatively defines a quantum field theory
(QFT), all numerical studies of QFT effectively fall in this second category of
applications \cite{Creutz-book, MM-book, Rothe-book, DGDT-book, GL-book}.

Since double (or higher) precision has become commonplace, it seems that little
concern has been shown about the possibility that machine precision limitations
might significantly influence the numerical results of MCMC simulations. Some
theoretical results regarding the behavior of the MCMC algorithm in the
presence of truncation errors have been obtained in Refs.~\cite{RRS-98,
BRS-01}, which, loosely speaking, state that the MCMC algorithm is stable against 
this type of perturbation. This means that the perturbed Markov chain still
converges to a unique invariant probability distribution, and this probability
distribution is ``close'' to that of the unperturbed Markov chain.  Similar
conclusions have been reached for stochastic perturbations of the MCMC
algorithm, the so called Markov chains in random environment, which have also
been considered in the mathematical literature, see, e.~g., \cite{Takahashi-69,
Cogburn-84}. Stochastic perturbations are relevant both for quantum computing
applications \cite{Temme:2009wa, Yung:2012jjy, QuBiPF:2020iiz, Aiudi:2023cyq,
Ballini:2023ljs} and for rounding errors in architectures for which thread
scheduling is non-deterministic.

These theoretical results show that the MCMC approach is robust with respect to
a large class of perturbations as long as we consider the first class of
applications introduced above, however they seems to be of little use to
understand whether universal properties are stable or not against perturbations
of the MCMC algorithm. Since continuous phase transitions correspond to points
of the phase diagram where the free energy is non-analytic in the thermodynamic
limit, the closeness of the invariant probability distribution function of the
perturbed MCMC to that of the exact MCMC, for fixed volume, is not enough to
guarantee that the two Markov chains display the same critical behavior.

A simple example can be useful to clearly understand this point:
if we consider the perturbation of the MCMC used to sample an Ising model
obtained by introducing an external magnetic field $h$, it is immediate to see
that, at fixed lattice volume, the probability distribution function of the
$h\neq 0$ model is arbitrarily close to that of the $h=0$ model as far as we
keep $|h|$ small enough. On the other hand, for whatever small but
non-vanishing value of $|h|$ there is no phase transition if we keep $h$ fixed
and perform the thermodynamic limit, although very large lattices may be needed
to notice this fact in an actual numerical simulation.

The aim of the present paper is to systematically investigate, by means of
numerical simulations, the effect that perturbations of the MCMC algorithm have
on the critical behavior of a classical statistical model.  In particular we
will use the three dimensional XY model as test bed (see, e.~g.,
\cite{Ballesteros:1996bd, Guida:1998bx, BMPS-06, Campostrini:2006ms,
Kompaniets:2017yct, Hasenbusch-19, Xu:2019mvy, Chester:2019ifh,
DePolsi:2020pjk} for some determinations of its critical properties). 

We consider locally reversible updates, i.~e. satisfying the detailed balance
principle, since these are so far the most commonly adopted ones, see however
\cite{Krauth-21} for a recent review on non-reversible update schemes.  Even
assuming reversibility there is however still much freedom in the
implementation of a MCMC algorithm to sample the configuration space of a given
classical statistical system, and only in few cases it is possible to associate
the Monte Carlo evolution to a physical real-time evolution (see, e.~g.,
\cite{HH-77, FM-06} for the corresponding critical dynamics). It is thus
possible that some implementations turn out to be more robust than others.  For
this reason we study both perturbations of the single site Metropolis update
\cite{MRRTT-53} and of the single cluster update \cite{Wolff-89},
considering deterministic and stochastic perturbations in both the cases. It
would obviously be possible to add also a microcanonical update, thus obtaining
an overrelaxed update scheme \cite{Adler-81}, however we prefer to consider
just the simplest local or cluster updates in order to simplify the
interpretation of the numerical results.

The paper is organized as follows: in Sec.~\ref{model} we introduce the MCMC
algorithms used to sample the configurations of the three dimensional XY model,
and describe the perturbations of these algorithms considered in the present
work.  In Sec.~\ref{numres} we define the main observables used to investigate
the critical behavior and recall some fundamental facts about Finite Size
Scaling (FSS), then we discuss the numerical results obtained using the
perturbed MCMC algorithms. Finally, in Sec~\ref{concl}, we draw our
conclusions.

\section{The XY model and the MCMC perturbations studied}
\label{model}

As anticipated in the introduction, we use the three dimensional XY model as
test bed to investigate how perturbations of the MCMC algorithm affect the
critical behavior of a classical spin system. The three dimensional XY model is
obtained by associating a two dimensional vector ${\bm s}_{\bm n}$, normalized
in such a way that $|\bm{s}_{\bm n}|=1$, to each site ${\bm n}$ of a three
dimensional lattice. The Hamiltonian of the system is
\begin{equation}
H[\{{\bm s}_{\bm n}\}]=-J\sum_{\langle {\bm n}, {\bm k}\rangle}{\bm s}_{\bm n}\cdot {\bm s}_{\bm k}\ ,
\end{equation}
where the notation $\langle \phantom{n},\phantom{k}\rangle$ is used to specify
that the sum extends on all neighboring sites of the lattice, and the partition
function is thus given by ($\beta=1/(k_BT)$)
\begin{equation}
Z=\sum_{ \{{\bm s}_{\bm n}\}} e^{-\beta H[\{{\bm s}_{\bm n}\}]}\ .
\end{equation}
We use the ferromagnetic XY model ($J>0$), and by measuring the coupling $J$ in
units of the temperature we can effectively set $J=1$ in the Hamiltonian. As
usual we perform simulations using cubic lattices with periodic boundary
conditions, and we denote by $L$ the linear extent of the lattice.

We consider two different update schemes: the single site Metropolis update and
the single cluster update. In the single site Metropolis update we sweep
through the lattice in lexicographic order, we generate a new random vector
${\bm s}^{(\mathrm{trial})}$ by performing a random rotation of
the spin ${\bm s}_{\bm n}$ (see later for more details), and accept the update
${\bm s}_{\bm n}\to {\bm s}^{(\mathrm{trial})}$ with probability 
\begin{equation}\label{eq:metro}
P=\mathrm{min}(1,e^{-\beta\Delta E})\ ,
\end{equation}
where $\Delta E=H(\{{\bm s}_{\bm n}\to {\bm s}^{(\mathrm{trial})}\})-H(\{{\bm
s}_{\bm n}\})$ is the energy change of the configuration induced by the
substitution ${\bm s}_{\bm n}\to {\bm s}^{(\mathrm{trial})}$. To
avoid the use of slow trigonometric functions the random rotation of the spin
is performed by using the auxilliary functions 
\begin{equation}
c(\theta)=\frac{1}{\sqrt{1+\theta^2}}\ ,\quad s(\theta)=\theta c(\theta)\ , 
\end{equation}
with $\theta$ uniformly distributed in  $(-2\pi, 2\pi)$, and 
\begin{equation}
\begin{aligned}
s^{\mathrm{trial}}_x=c(\theta) ({\bm s}_{\bm n})_x - s(\theta) ({\bm s}_{\bm n})_y \ ,\\
s^{\mathrm{trial}}_y=s(\theta) ({\bm s}_{\bm n})_x + c(\theta) ({\bm s}_{\bm n})_y \ .
\end{aligned}
\end{equation}

 In the single
cluster update \cite{Wolff-89} we instead randomly select the
direction ${\bm v}$ ($|{\bm v}|=1$) in the spins two dimensional space, and
we denote by $R_{\bm v}({\bm s})$ the reflection with respect to plane
orthogonal to ${\bm v}$:
\begin{equation}
R_{\bm v}({\bm s})={\bm s} -2({\bm v}\cdot{\bm s}){\bm v}\ .
\end{equation}
The real update starts by randomly picking a site ${\bm n}$ and performing the
substitution ${\bm s}_{\bm n} \to R_{\bm v}({\bm s}_{\bm n})$. The site ${\bm
n}$ is then used as a seed to build a cluster, by recursively adding neighboring
sites with probability 
\begin{equation}\label{eq:cluster}
P({\bm s}_{\bm i}, {\bm s}_{\bm o})= 
1-\exp\Big(\mathrm{min}\big[0, 
2 \beta ({\bm s}_{\bm i}\cdot {\bm v})({\bm s}_{\bm o}\cdot{\bm v})\big]\Big)\ ,
\end{equation}
where ${\bm s}_{\bm i}$ is a spin already included in the cluster and ${\bm
s}_{\bm o}$ is a next-neighbor of the spin ${\bm s}_{\bm i}$ not yet included
in the cluster. The spin of any site added to the cluster is updated using the
reflection $R_{\bm v}$, and the construction goes on until an iteration is
reached in which no further sites are added to the cluster.

With the purpose of investigating the stability of the XY critical properties
against perturbations of the MCMC algorithm, we now introduce some deterministic
and stochastic perturbations of the single site Metropolis and of the single
cluster update schemes. Both the original algorithms (used to estimate some
universal FSS curves of the XY universality class, see Sec.~\ref{numres}) and
the perturbed ones have been implemented in \texttt{C} language using double
precision arithmetic. 

Let us start from the perturbations of the single site Metropolis update. The
essential step of this algorithm is the accept/reject step (sometimes called
Metropolis filter) in Eq.~\eqref{eq:metro}, so we consider perturbations
corresponding to ``wrong'' estimates of the energy difference $\Delta E$.  A
simple stochastic perturbation is obtained by considering in
Eq.~\eqref{eq:metro} the substitution
\begin{equation}\label{eq:metro_rand}
\Delta E\to \Delta E+G'_{\sigma}-G_{\sigma}\ ,
\end{equation}
where $G'_{\sigma}$ and $G_{\sigma}$ are two independent Gaussian random
variables with zero average and variance $\sigma^2$. This perturbation can be
thought of as a simple modeling of the case in which the energies
of the initial and trial states are not exactly computed, but stochastically
estimated.  A different (not used) possibility would be to add to $\Delta E$ a
single Gaussian random variable, to model the case in which directly $\Delta E$
is stochastically estimated. To mimic the effect of a finite machine precision
we can instead use
\begin{equation}\label{eq:metro_cut}
\Delta E \to \mathrm{round}(\Delta E\cdot\Lambda)/\Lambda\ ,
\end{equation}
where $\mathrm{round}(x)$ denotes the rounding to the closest integer of $x$,
which amounts to truncate $\Delta E$ with an accuracy $1/\Lambda$. To provide a
scale for the values of $\sigma$ and $\Lambda$ used in the following section
we note that the average value of $\Delta E$ at the critical point of the XY
model is $\approx 2$.

In the case of the single cluster update we modify the probability of adding a
new site to the cluster, see Eq.~\eqref{eq:cluster}. A random perturbation is
obtained by performing the substitution
\begin{equation}\label{eq:cluster_rand}
({\bm s}_{\bm i}\cdot {\bm v})({\bm s}_{\bm o}\cdot{\bm v})\to
({\bm s}_{\bm i}\cdot {\bm v})({\bm s}_{\bm o}\cdot{\bm v})+G_{\sigma}\ ,
\end{equation}
where once again $G_{\sigma}$ is a random Gaussian variable with zero average
and variance $\sigma^2$. A finite precision truncation is instead obtained by using the
substitution
\begin{equation}\label{eq:cluster_cut}
({\bm s}_{\bm i}\cdot {\bm v})({\bm s}_{\bm o}\cdot{\bm v})\to
\mathrm{round}\Big[({\bm s}_{\bm i}\cdot {\bm v})({\bm s}_{\bm o}\cdot{\bm v})\Lambda\Big]/\Lambda\ ,
\end{equation}
where $\mathrm{round}$ has been defined below Eq.~\eqref{eq:metro_cut}. At the
transition point of the XY model the average value of the quantity  $({\bm
s}_{\bm i}\cdot {\bm v})({\bm s}_{\bm o}\cdot{\bm v})$ is $\approx 0.25$.  

Let us stress that the mere fact that the typical value of the control
parameter is smaller for the single cluster update than for the single site
Metropolis update does not imply that the local update is more robust than the
cluster one. The two update algorithms are so different from each other that it
is not possible to draw such a conclusion based on this fact alone.

\section{Numerical results}
\label{numres}

\subsection{Observables and Finite Size Scaling}
\label{sec:FSS}

The observables we study can be defined starting from the two point correlation
function of the spin:
\begin{equation}
G({\bm x}-{\bm y})=\langle {\bm s}_{\bm x}\cdot {\bm s}_{\bm y}\rangle\ .
\end{equation}
In particular we consider Renormalization Group (RG) invariant observables,
since the form of their FSS is particularly simple and allows for easy
comparison between different models. The two RG invariant quantities considered
are the Binder cumulant
\begin{equation}
U=\frac{\langle \mu_2^2\rangle}{\langle\mu_2\rangle^2}\ ,\quad
\mu_2=\frac{1}{L^6}\sum_{{\bm x},{\bm y}}{\bm s}_{\bm x}\cdot{\bm s}_{\bm y}\ ,
\end{equation}
and the ratio
\begin{equation}
R_{\xi}=\xi/L\ ,
\end{equation}
where the second moment correlation length $\xi$ is defined by 
\begin{equation}
\xi^2=\frac{1}{4\sin^2(p_{\mathrm{min}}/2)}
\frac{\widetilde{G}({\bm 0})-\widetilde{G}({\bm p})}{\widetilde{G}({\bm p})}\ .
\end{equation}
In this expression $p_{\mathrm{min}}=2\pi/L$ is the minimum non-vanishing
momentum consistent with periodic boundary conditions, ${\bm
p}=(p_{\mathrm{min}},0,0)$, and $\widetilde{G}({\bm k})$ is the Fourier
transform of the two point correlation function $G({\bm x})$ computed at
momentum ${\bm k}$.

\begin{table}[t]
\begin{tabular}{l|l}
\hline\hline
$\beta_c$   & 0.45416474(10)[7] \\ \hline
$\nu$       & 0.67169(7) \\ \hline
$\omega$    & 0.789(4) \\ \hline
\rule[-1.5mm]{0mm}{4.5mm}$R_{\xi}^*$ & 0.59238(7) \\ \hline
$U^*$       & 1.24296(8) \\ \hline\hline
\end{tabular}
\caption{Critical temperature and (some) universal critical properties of the XY
model, from \cite{Hasenbusch-19} (see, e.g.,  \cite{Ballesteros:1996bd, Guida:1998bx,
BMPS-06, Campostrini:2006ms, Kompaniets:2017yct, Xu:2019mvy, Chester:2019ifh,
DePolsi:2020pjk} for other determinations).}
\label{tab:XY}
\end{table}

If we introduce the standard notation $\tau=(\beta-\beta_c)/\beta_c$, the
FSS of a generic RG invariant quantity $R$ can be written in the form
(see, e.~g., \cite{PV-02})
\begin{equation}\label{eq:FSS_RG}
R(\beta, L)\simeq \mathcal{R}(L^{1/\nu}\tau)+L^{-\omega}\mathcal{R}_{\omega}(L^{1/\nu}\tau)+\cdots\ .
\end{equation}
In this expression $\nu$ is the critical exponent related to the divergence at
$\beta=\beta_c$ of the correlation length in the thermodynamic limit,
$\omega>0$ is (up to a sign) the scaling dimension of the first irrelevant
operator at the transition, and the final dots stand for further irrelevant
contributions and analytic background terms. $\mathcal{R}$ is a scaling
function universal up to a rescaling of its argument, while
$\mathcal{R}_{\omega}$ is universal up to a multiplicative constant and up to a
rescaling of its argument, as can be seen starting from the standard scaling
form of free energy density (see, e.g., \cite{PV-02}). In particular it follows
that
\begin{equation}
\lim_{L\to\infty}R(\beta_c,L)=\mathcal{R}(0)
\end{equation} 
is universal.  For the convenience of the reader we report in Tab.~\ref{tab:XY}
the universal values the observables $R_{\xi}$ and $U$ (denoted by an
asterisk) and of the critical exponents $\nu$ and $\omega$ for the three
dimensional XY universality class, as well as the value of the critical
temperature $\beta_c$ for the XY model.

Since $R_{\xi}$ turns out to be a monotonically increasing function of $\beta$
(and thus of $\tau$), it is possible to invert the relation $\tau\to R_{\xi}$
and rewrite the FSS for $U$ as a function of $R_{\xi}$: 
\begin{equation}\label{FSS_RG2}
U(\beta, L)=\mathcal{U}(R_{\xi})+L^{-\omega}\mathcal{U}_{\omega}(R_{\xi})+\cdots\ .
\end{equation}
In this expression $\mathcal{U}$ is a universal scaling function independent
of any non-universal rescaling factor, while $\mathcal{U}_{\omega}$ is still
universal only up to a multiplicative constant. Verifying that Eq.~\eqref{FSS_RG2}
holds with the same asymptotic scaling curve $\mathcal{U}(R_{\xi})$ is thus a
particularly convenient way of testing whether two different models are in the
same universality class. 

\begin{table}
\begin{tabular}{l|l}
\hline\hline
$b_0$ &             -17.46924088170859 \\ \hline  
$b_1$ &   \phantom{-1}0.66215088251974 \\ \hline
$b_2$ &   \phantom{1}-0.82048697561210 \\ \hline
$b_3$ &   \phantom{-}13.78704076975388 \\ \hline
$b_4$ &   \phantom{-1}3.85932964105796 \\ \hline
$b_5$ &   \phantom{1}-1.29662539205099 \\ \hline \hline
\end{tabular}
\begin{tabular}{l|l}
\hline\hline
$c_0$ &   \phantom{-}10.70348785915293 \\ \hline
$c_1$ &   \phantom{-1}0.37099210816252 \\ \hline
$c_2$ &   \phantom{1}-0.42383845323376 \\ \hline
$c_3$ &             -10.06615359507961 \\ \hline
$c_4$ &   \phantom{1}-1.31489275164629 \\ \hline
$c_5$ &   \phantom{-1}0.59437782721002 \\ \hline \hline
\end{tabular}
\caption{Optimal fit parameters obtained by fitting data for $U(R_{\xi})$ of
the XY model in the interval $R_{\xi}\in [0.3,1]$ using the parametrization in
Eqs.~\eqref{eq:fit_func}-\eqref{eq:fit_func2}. To perform the fit we only
consider lattices with $L\ge 16$.}
\label{tab:fit_param}
\end{table}

\subsection{XY scaling curve}
\label{sec:XYscal}

Since the function $\mathcal{U}(R_{\xi})$ entering Eq.~\eqref{FSS_RG2} is
universal, it will be useful in the following to have a parametric
representation of it, to investigate whether the results obtained using the
perturbed models display the same critical behavior of the original XY model. 

\begin{figure}[t]
\includegraphics[width=0.9\columnwidth, clip]{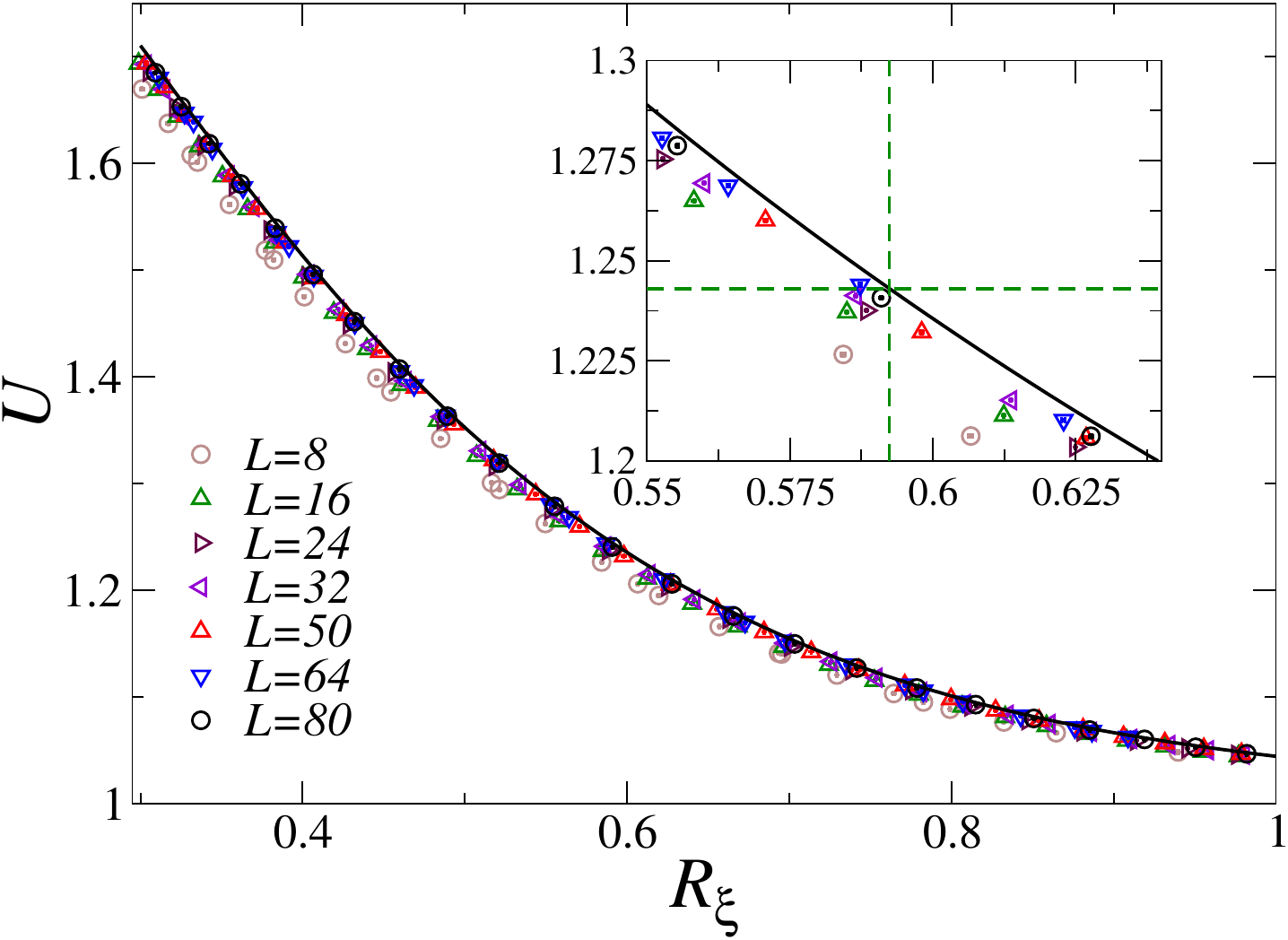}\\
\includegraphics[width=0.9\columnwidth, clip]{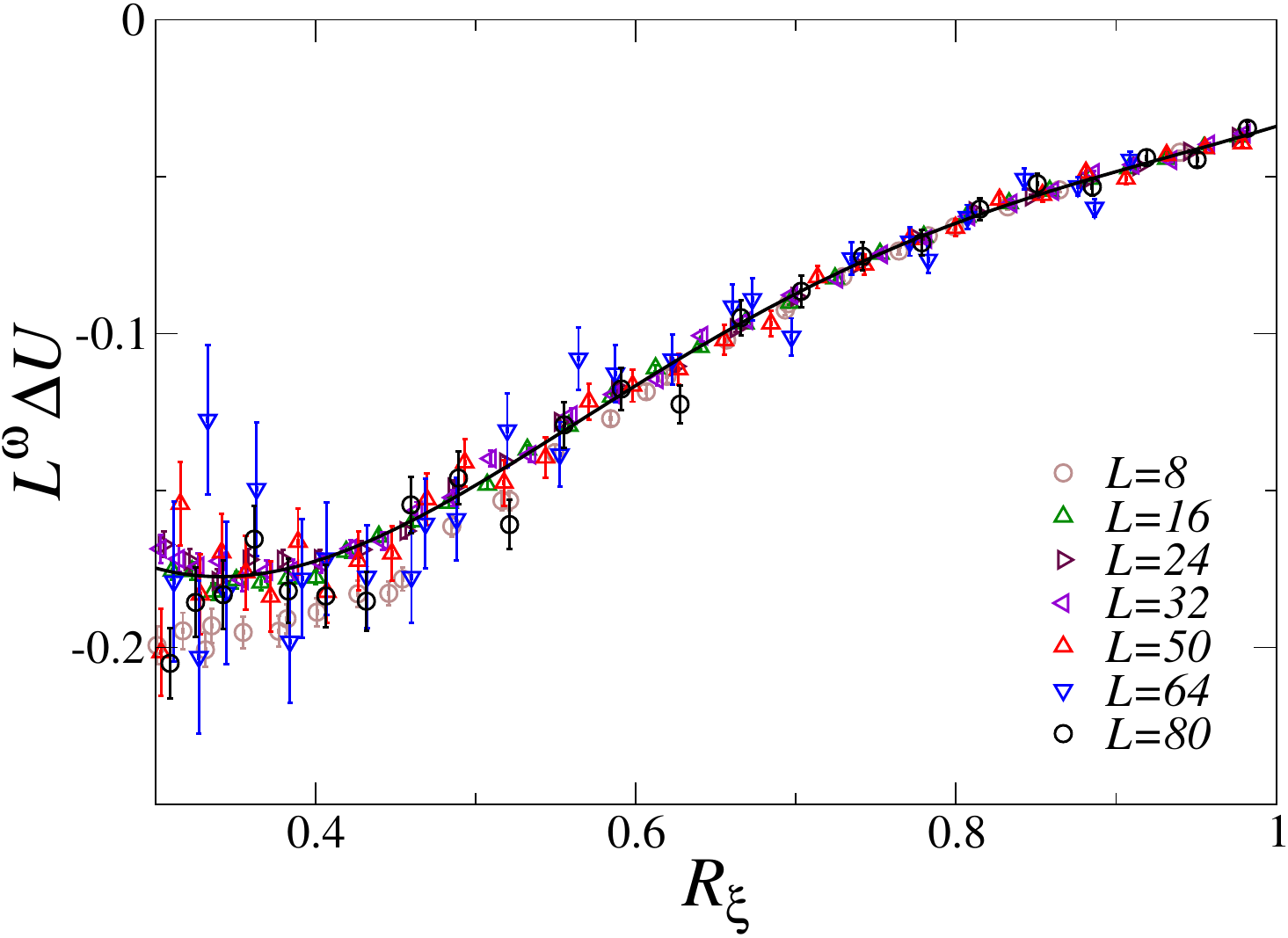}
\caption{
Numerical estimates of the scaling curves $\mathcal{U}(R_{\xi})$ and
$\mathcal{U}_{\omega}(R_{\xi})$ (see Eq.~\eqref{FSS_RG2}) of the XY model.
(upper panel) Numerical results for $U$ as a function of $R_{\xi}$. The solid
line is the result of the fit to Eqs.~\eqref{eq:fit_func}-\eqref{eq:fit_func2}
described in the main text. The inset shows a zoom of the region corresponding
to $\beta\approx \beta_c$, and dashed lines are drawn in correspondence of the
known universal values $R_{\xi}^*$ and $U^*$, see Tab.~\ref{tab:XY}, which have
not been used in the fitting procedure. (lower panel) Numerical results for
$L^{\omega}(U(R_{\xi})-\mathcal{U}(R_{\xi}))\approx\mathcal{U}_{\omega}(R_{\xi})$,
computed by using the value of the leading correction exponent reported in
Tab.~\ref{tab:XY}.  The solid line is the result of the fit to
Eqs.~\eqref{eq:fit_func}-\eqref{eq:fit_func2} described in the main text. 
}
\label{fig:fit_XY}
\end{figure}

To parametrize $\mathcal{U}(R_{\xi})$ in the XY model we use the same
functional form already adopted in
\cite{Bonati:2021hzo}
\begin{equation}\label{eq:fit_func}
\begin{aligned}
f(x)=&2+a_0x^2(1-e^{-8x^2})-\\
&-(a_1+a_2x)(1-e^{-7x^2})+\sum_{i=3}^5 a_ix^{i-1}\ ,
\end{aligned}
\end{equation}
where the coefficients $a_i$ have a $L$ dependence which takes into account the
leading scaling corrections
\begin{equation}\label{eq:fit_func2}
a_k=b_k+L^{-\omega}c_k\ .
\end{equation}

We simulated several values of $\beta$ for $L=8$, 16, 24, 32, 50, 64, and 80,
and performed the fit to Eq.~\eqref{eq:fit_func} on the interval $0.3\le
R_{\xi}\le 1$, using only data coming from lattices of linear extent $L\ge 16$.
The optimal values obtained for the fit parameters are reported in
Tab.~\ref{tab:fit_param}, and using a bootstrap procedure we estimated the
average relative error of $\mathcal{U}(R_{\xi})$ to be $\approx 8\times
10^{-5}$ (with the maximum relative error being $\approx 2\times 10^{-4}$), and
the average relative error of $\mathcal{U}_{\omega}(R_{\xi})$ to be $\approx
0.013$ (with the maximum relative error being $\approx 0.046$). Note that these
values are just statistical errors, and do not take into account the
systematics related to the choice of the fit function. We will come back to
this point in a moment. 

Numerical data for $U$ as a function of $R_{\xi}$ are shown in
Fig.~\ref{fig:fit_XY}, together with the estimates of $\mathcal{U}(R_{\xi})$
and $\mathcal{U}_{\omega}(R_{\xi})$ obtained by the above described fit.  In
the upper panel of Fig.~\ref{fig:fit_XY} a zoom is shown of the region
corresponding to $\beta\approx \beta_c$, with dashed lines corresponding to the
known values of the universal quantities $R_{\xi}^*$ and $U^*$ (see
Tab.~\ref{tab:XY}). Using $R_{\xi}^*=0.59238$ we obtain for $U^*$ the estimate
$f(R^*_{\xi})= 1.24306(11)$ which is nicely consistent with the known value
1.24296(8) obtained in Ref.~\cite{Hasenbusch-19} and reported in
Tab.~\ref{tab:XY}. This is a good indication that systematic errors related to
the fitting procedure are likely smaller than the statistical ones, at least
for $\mathcal{U}(R_{\xi})$ and $R_{\xi}\approx R_{\xi}^*$. 

\begin{table}[t]
\begin{tabular}{lll}
\hline\hline
$\sigma^2$ & $\beta_c(R)$   & $\beta_c(U)$ \\ \hline
0.01       & 0.4548716(65)  & 0.454884(15) \\ \hline
0.1        & 0.4613098(46)  & 0.461329(19) \\ \hline
1          & 0.533932(11)   & 0.533967(19) \\ \hline\hline
\end{tabular}
\caption{Single site Metropolis update with Gaussian random energy perturbation (see
Eq.~\eqref{eq:metro_rand}): estimates of the critical temperature for different
values of $\sigma^2$, obtained by analyzing $R_{\xi}$ or $U$. For comparison
the critical temperature of the XY model is 0.45416474(10)[7] (from
Ref.~\cite{Hasenbusch-19}).
Note that the critical temperature is not universal, and there is
no reason to expect $\beta_c$ not to vary with $\sigma^2$. Its variation is in
fact an indication that the values of $\sigma^2$ used are large enough to have an effect on the system.
}
\label{tab:betac_metro_rand}
\end{table}

\begin{table}[t]
\begin{tabular}{lll}
\hline\hline
$\Lambda$ & $\beta_c(R)$   & $\beta_c(U)$ \\ \hline
$10^3$    & 0.4542428(83)  & 0.4542441(55) \\ \hline
10        & 0.4621358(86)  & 0.462136(13)  \\ \hline
2         & 0.498336(10)   & 0.498350(15)  \\ \hline\hline
\end{tabular}
\caption{Single site Metropolis update with energy truncation perturbation (see
Eq.~\eqref{eq:metro_cut}): estimates of the critical temperature for different
values of $\Lambda$, obtained by analyzing $R_{\xi}$ or $U$. For comparison
the critical temperature of the XY model is 0.45416474(10)[7] (from
Ref.~\cite{Hasenbusch-19}).
Note that the critical temperature is not universal, and there is
no reason to expect $\beta_c$ not to vary with $\Lambda$. Its variation is in
fact an indication that the values of $\Lambda$ used are large enough to have an effect on the system.
}
\label{tab:betac_metro_cut}
\end{table}

In the lower panel of Fig.~\ref{fig:fit_XY} we also show results related to the
numerical estimate of $\mathcal{U}_{\omega}(R_{\xi})$, from which it can be
seen that data obtained from the $L=8$ lattice in these
high-accuracy simulations systematically deviate from the other ones; this is
why we included only $L\ge 16$ data in the fit.  For what concerns the
systematic errors of the $\mathcal{U}_{\omega}(R_{\xi})$ determination, we
expect them to be larger than those affecting $\mathcal{U}(R_{\xi})$, but we
do not have data to quantify them in a robust way.  For this reason, in the
following analyses of the perturbed models, we will just use
$\mathcal{U}(R_{\xi})$, and in particular the fact that
$(U-\mathcal{U}(R_{\xi}))L^{\omega}$ converges to a scaling curve, without
explicitly comparing this scaling curve with the estimate of
$\mathcal{U}_{\omega}(R_{\xi})$ obtained from the XY model.  Of course we could
also directly compare data reported in the lower panel of Fig.~\ref{fig:fit_XY}
with the corresponding data of the perturbed model, but we would anyway have to
fix the value of a non-universal parameter, see Sec.~\ref{sec:FSS}. Since
however statistical errors of the quantity $(U-\mathcal{U}(R_{\xi}))L^{\omega}$
will turn out to be quite large for the perturbed models, the result of this
comparison (although positive) is not particularly significant, and will not be
reported in the following.

\subsection{Perturbed MCMC results}

\subsubsection{Single site Metropolis algorithm}

We now present the results obtained by using the perturbed single site
Metropolis update to sample the configurations of the three dimensional XY model.

\begin{figure}[b]
\includegraphics[width=0.9\columnwidth, clip]{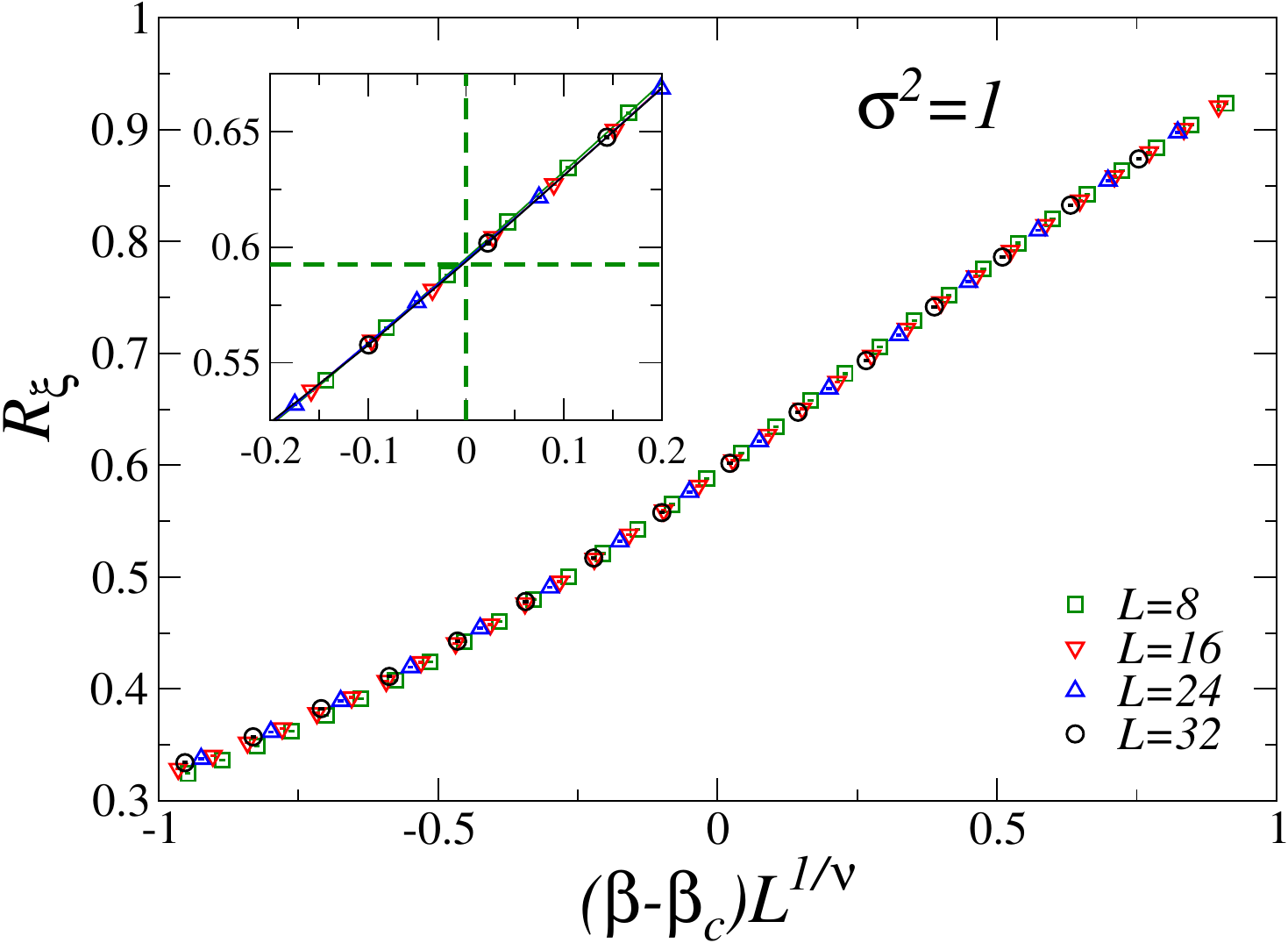}\\
\includegraphics[width=0.9\columnwidth, clip]{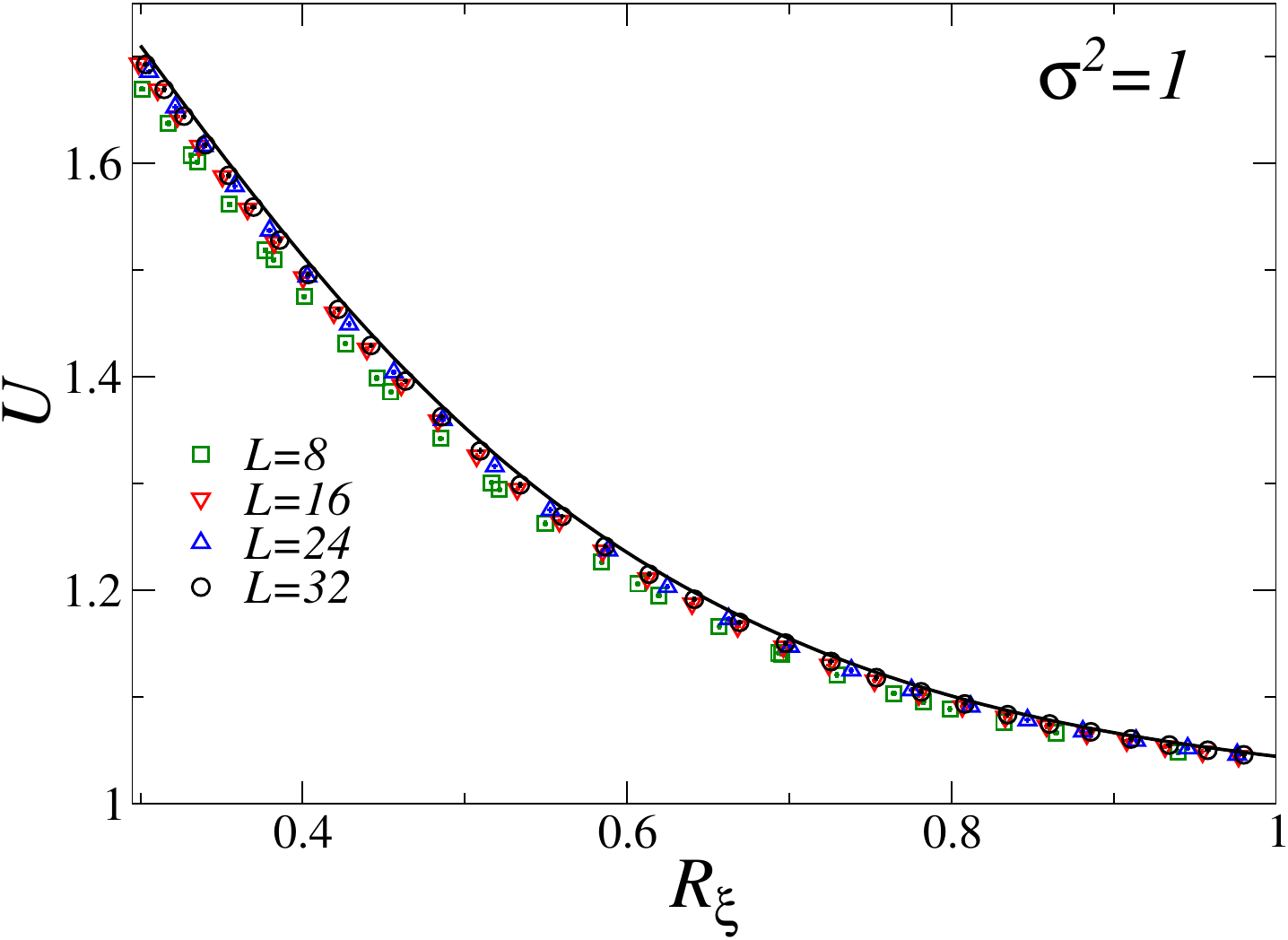}
\caption{Results for the single site Metropolis update with Gaussian random
energy perturbation (see Eq.~\eqref{eq:metro_rand}) with variance $\sigma^2=1$.
(upper panel) FSS of $R_{\xi}$ obtained by using the value of the critical
temperature reported in Tab.~\ref{tab:betac_metro_rand} and the value of the
critical exponent $\nu$ reported in Tab.~\ref{tab:XY}.  The inset shows a zoom
of the region close to $\beta\approx \beta_c$, and dashed lines are drawn in
correspondence of $\beta=\beta_c$ and $R_{\xi}=R_{\xi}^*$.  (lower panel)
Scaling of $U$ against $R_{\xi}$. The solid line is the parametrization of the
universal scaling curve $\mathcal{U}(R_{\xi})$ of the XY model obtained in
Sec.~\ref{sec:XYscal}.  }
\label{fig:metro_rand_sigma1_1}
\end{figure}

\begin{figure}[t]
\includegraphics[width=0.9\columnwidth, clip]{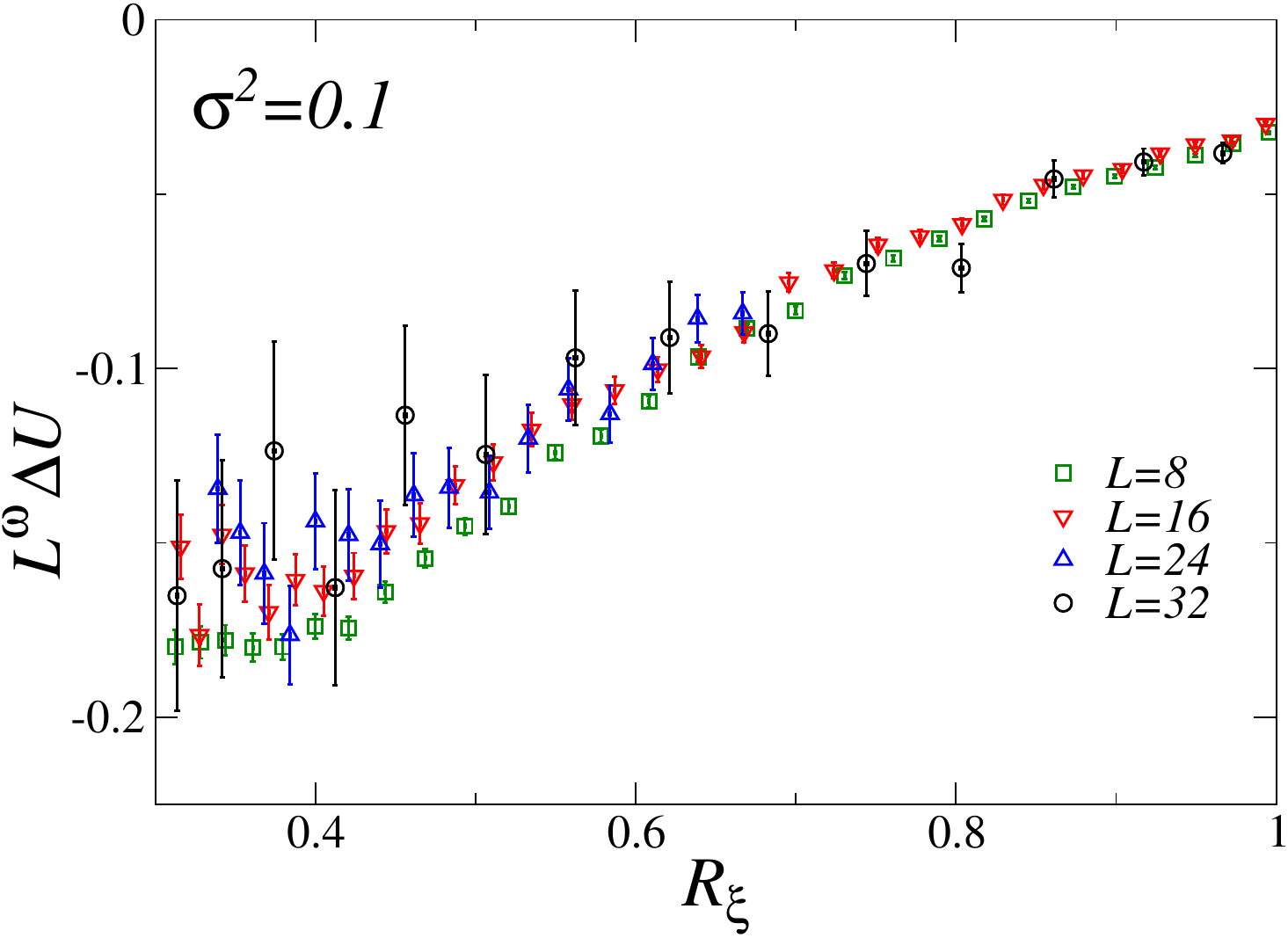}\\
\includegraphics[width=0.9\columnwidth, clip]{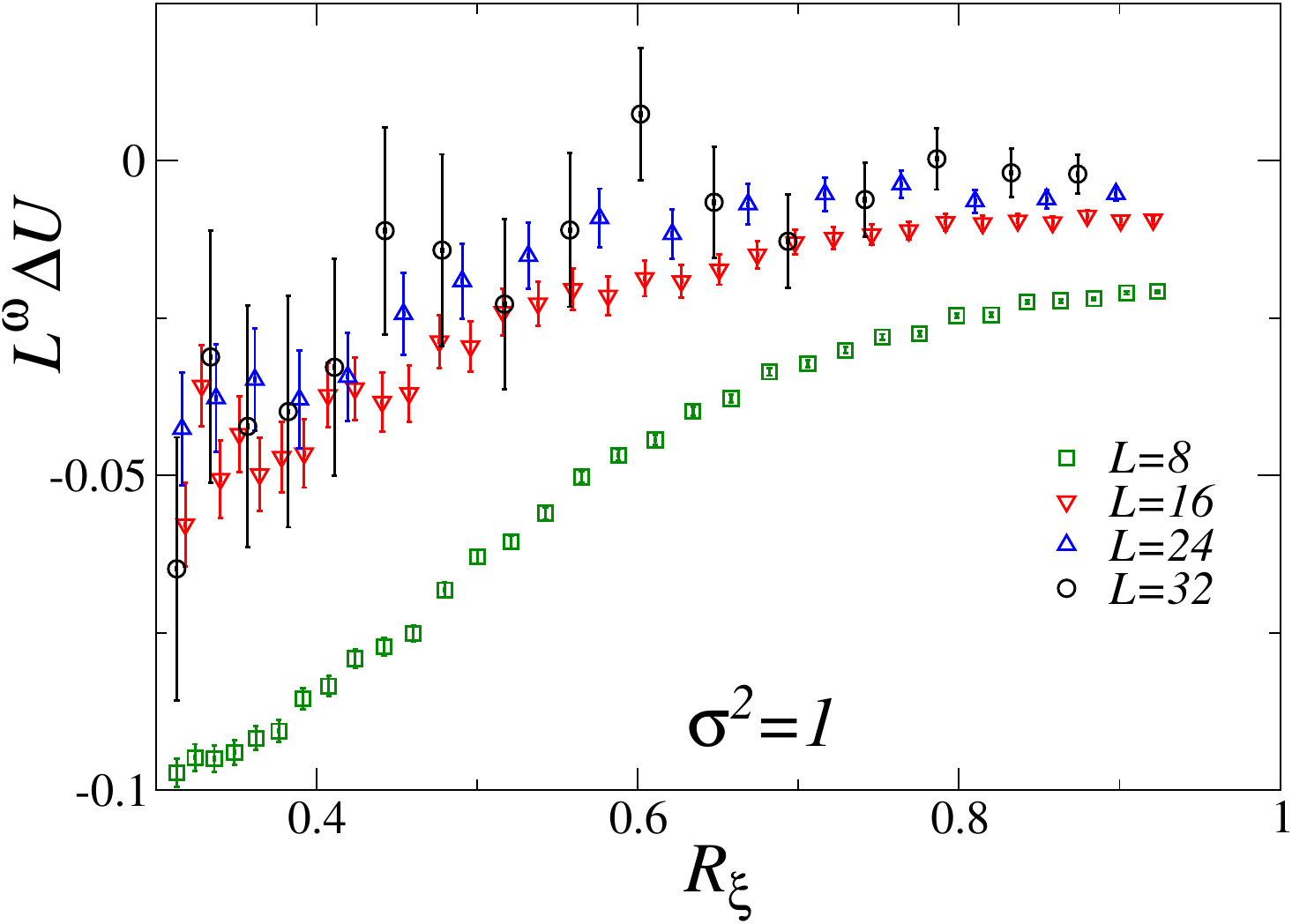}
\caption{Results for the single site Metropolis update with Gaussian random
energy perturbation (see Eq.~\eqref{eq:metro_rand}). Estimates of
$L^{\omega}(U(R_{\xi})-\mathcal{U}(R_{\xi}))$, for the model with
$\sigma^2=0.1$ (upper panel) and $\sigma^2=1$ (lower panel).}
\label{fig:metro_rand_sigma1_2}
\end{figure}

Let us start from the stochastic perturbation described by
Eq.~\eqref{eq:metro_rand}, in which a Gaussian noise with zero average and
variance $\sigma^2$ is added to the energies of the configurations.  We
considered three values of the variance ($\sigma^2=0.01, 0.1, 1$), and for each value of $\sigma^2$
we performed simulation at several values of $\beta$ using four different
lattice sizes ($L=8, 16, 24, 32$). 

To quantify the effect of the perturbation on non-universal quantities, for
each value of $\sigma^2$ we estimate the critical temperature from the crossing
points of $R_{\xi}$ and $U$ as a function of $\beta$. More precisely, we use
the FSS reported in Eq.~\eqref{eq:FSS_RG}, with polynomial approximations of
the functions $\mathcal{R}(x)$ and $\mathcal{R}_{\omega}(x)$, to fit data. In
all the cases we fix $\nu$, $\omega$, $R_{\xi}^*$ and $U^*$ using the universal
values of the XY universality class reported Tab.~\ref{tab:XY}.  Already the
fact that the fit used to extract $\beta_c$ results in a
$\chi^2/\mathrm{dof}\approx 1$ is thus a first indication that universal
quantities are not affected by the stochastic noise.  The residual dependence
of the results on the degree of the polynomials used in the fit,
on the fit range adopted, and on the minimum value of $L$
considered in the fit, has been added as a systematic error to the statistical error, and
the final results are reported in Tab.~\ref{tab:betac_metro_rand}. The good
agreement between the estimates obtained using $R_{\xi}$ and those obtained
using $U$ is an indication that systematic errors have been correctly taken
into account.

\begin{figure}[t]
\includegraphics[width=0.9\columnwidth, clip]{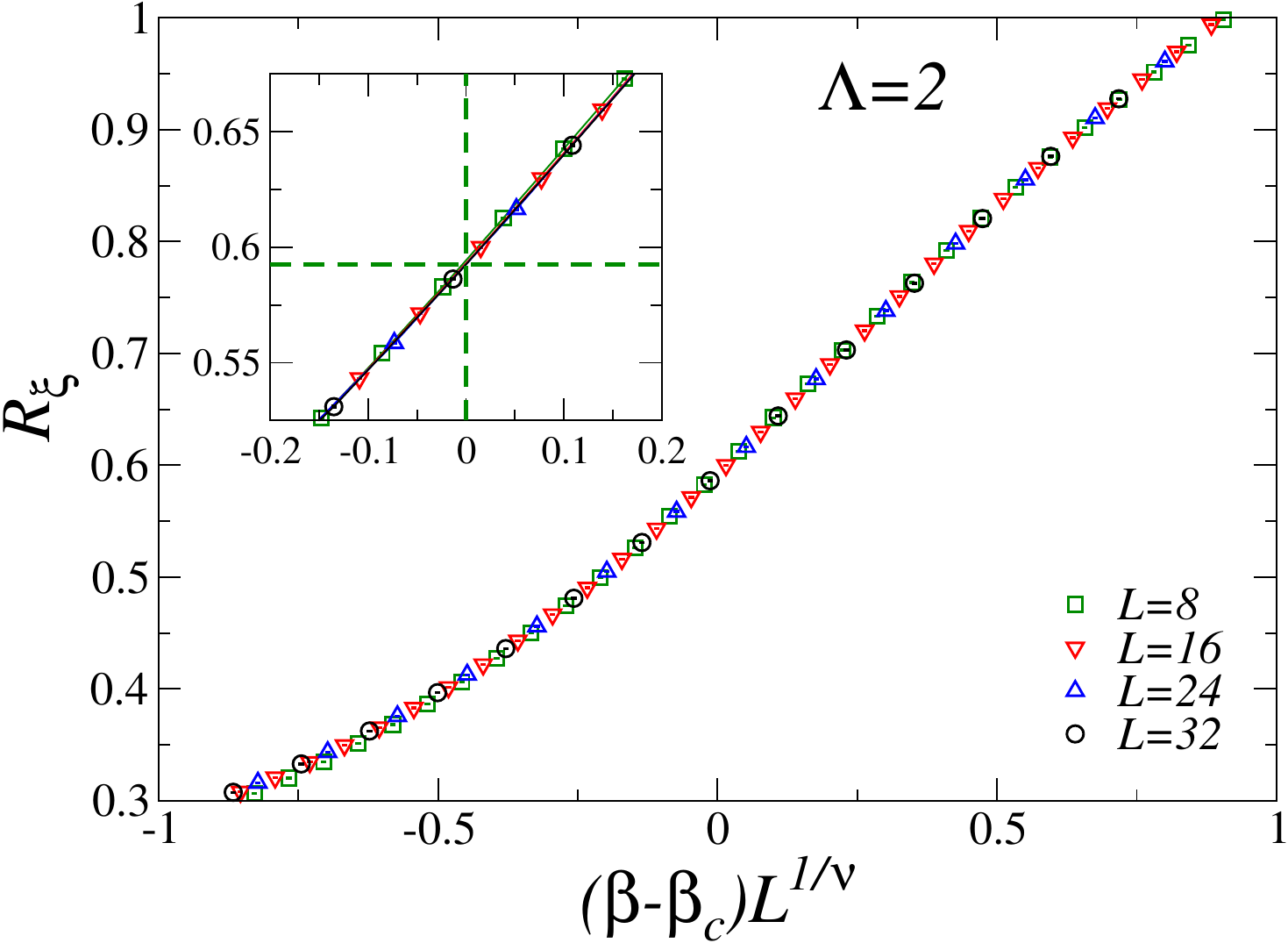}
\caption{Results for the single site Metropolis update with truncation energy
perturbation (see Eq.~\eqref{eq:metro_cut}) with truncation cut-off
$\Lambda=2$.  FSS of $R_{\xi}$ obtained by using the value of the critical
temperature reported in Tab.~\ref{tab:betac_metro_cut} and the value of the
critical exponent $\nu$ reported in Tab.~\ref{tab:XY}.  The inset shows a zoom
of the region close to $\beta\approx \beta_c$, and dashed lines are drawn in
correspondence of $\beta=\beta_c$ and $R_{\xi}=R_{\xi}^*$.  }
\label{fig:metro_cut_lambda2_1}
\end{figure}

\begin{figure}[h!]
\includegraphics[width=0.9\columnwidth, clip]{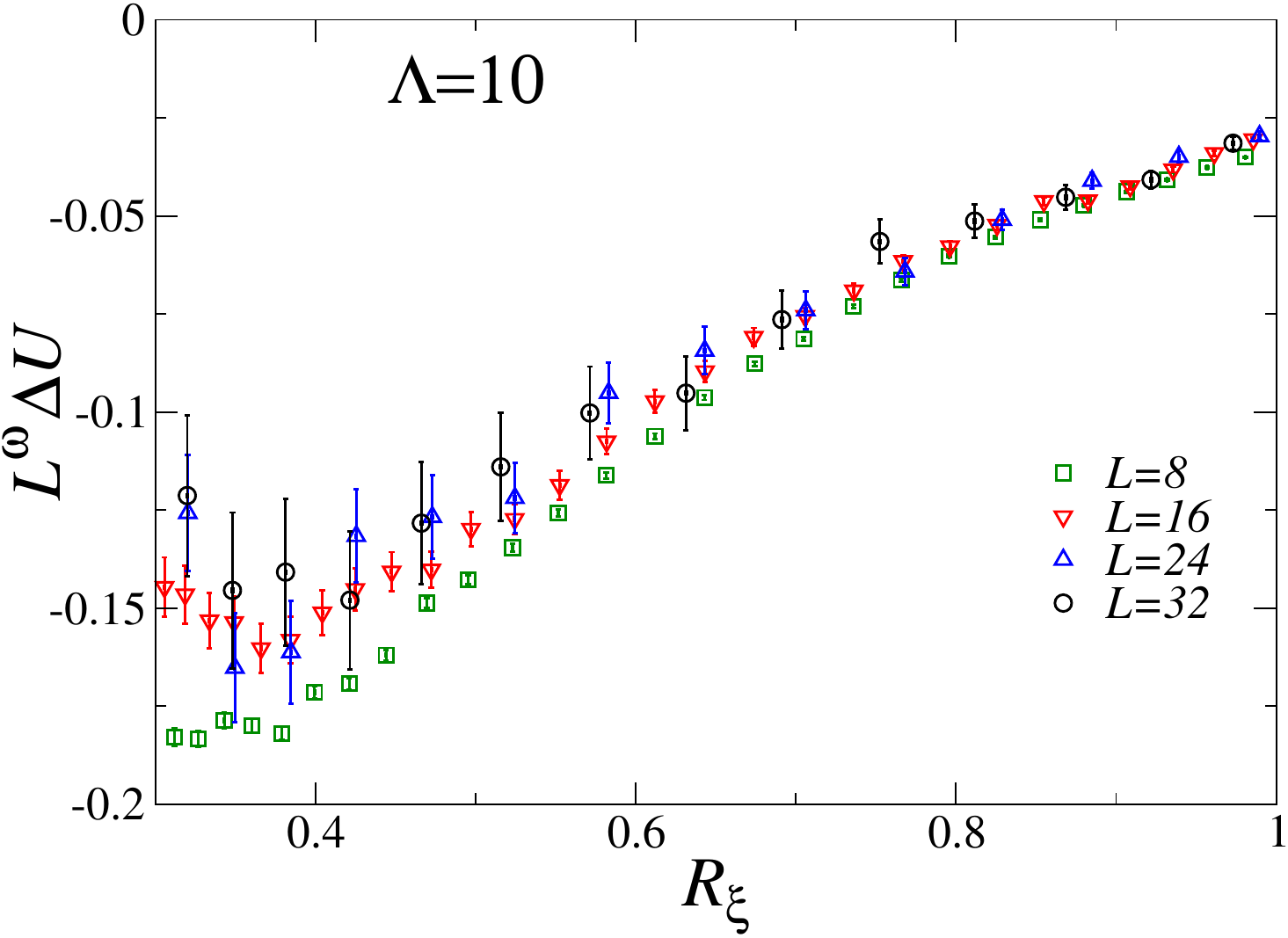}\\
\includegraphics[width=0.9\columnwidth, clip]{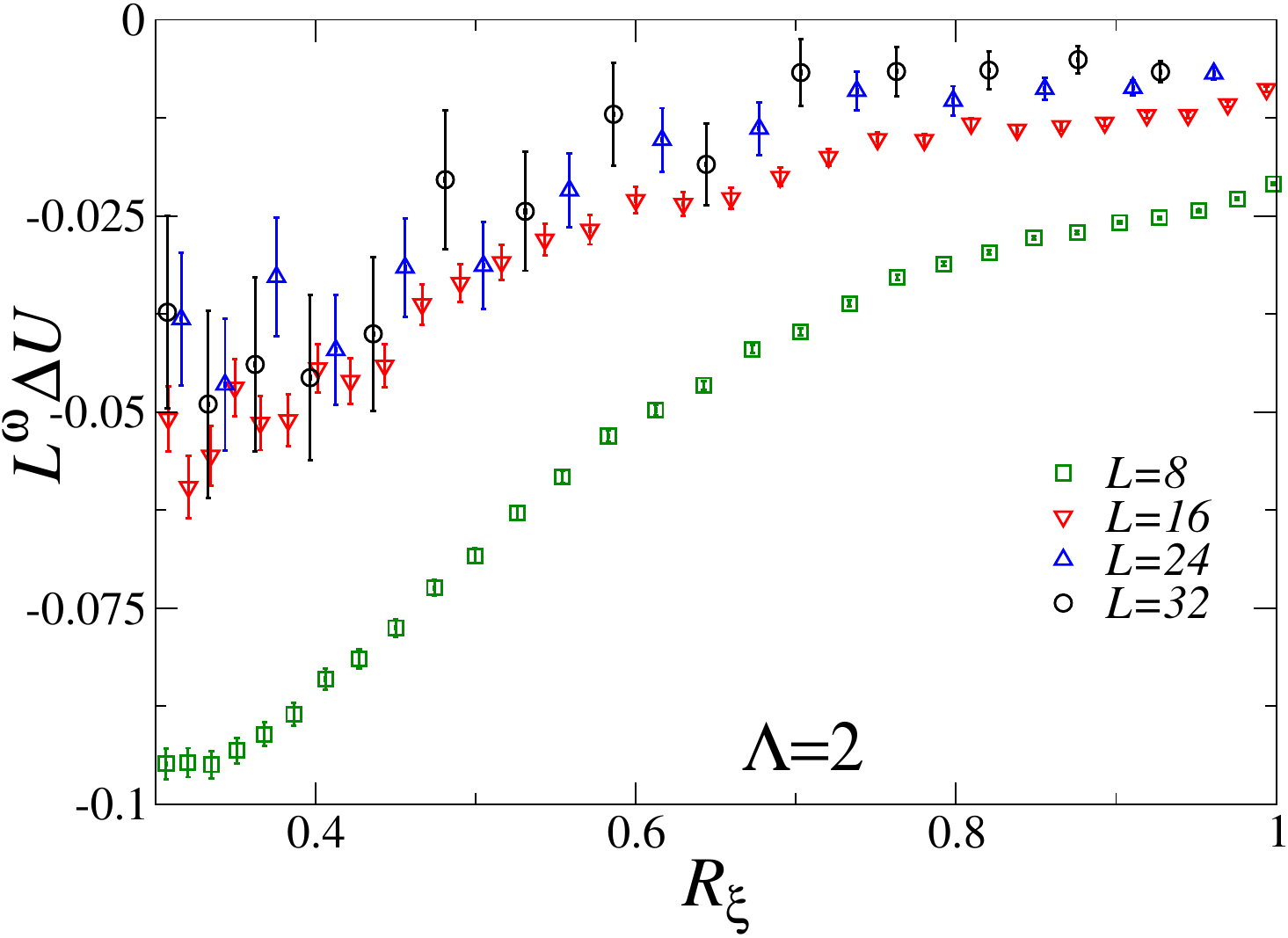}
\caption{Results for the single site Metropolis update with truncation energy
perturbation (see Eq.~\eqref{eq:metro_cut}). Estimates of
$L^{\omega}(U(R_{\xi})-\mathcal{U}(R_{\xi}))$,
for the model with $\Lambda=10$ (upper panel) and $\Lambda=2$ (lower panel).}
\label{fig:metro_cut_lambda2_2}
\end{figure}

We discuss in detail just the case of the largest perturbation considered
(i.~e. $\sigma^2=1$), since the conclusions are the same also for the other
values of $\sigma^2$ investigated.  In Fig.~\ref{fig:metro_rand_sigma1_1}
(upper panel) we show data for $R_{\xi}$ as a function of the scaling variable
$(\beta-\beta_c) L^{1/\nu}$, obtained by using the result reported in
Tab.~\ref{tab:betac_metro_rand} for $\beta_c$ and the known value of the XY
critical exponent $\nu$. The almost perfect scaling is a strong indication that
universal quantities are unaffected by this perturbation of the MCMC algorithm. 

The same conclusion can be reached from the scaling of $U$ as a function of
$(\beta-\beta_c) L^{1/\nu}$ (not shown), and also by looking at the scaling $U$
against $R_{\xi}$, shown in Fig.~\ref{fig:metro_rand_sigma1_1} (lower panel).
It is indeed quite clear that data are approaching the universal XY curve
$\mathcal{U}(R_{\xi})$ previously determined in Sec.~\ref{sec:XYscal}. To
better appreciate this fact we report in Fig.~\ref{fig:metro_rand_sigma1_2}
data for $L^{\omega}(U-\mathcal{U}(R_{\xi}))$, obtained by using the known XY
value of the leading correction to scaling exponent $\omega$ (see
Tab.~\ref{tab:XY}) and the universal XY curve $\mathcal{U}(R_{\xi})$.  The fact
that data asymptotically (i.~e. for large $L$) approach a common scaling curve
implies both that $\mathcal{U}(R_{\xi})$ correctly describes the asymptotic
behavior of $U$ as a function of $R_{\xi}$, and that $\omega=\omega_{XY}$ is
the leading scaling correction exponent. It is interesting to note that the
size of the leading corrections to scaling is actually reduced by increasing
the value of variance $\sigma^2$, as can be seen from
Fig.~\ref{fig:metro_rand_sigma1_2}, where data for $\sigma^2=0.1$ and
$\sigma^2=1$ are reported (note the different scale of the vertical axis in the
two panels). This could explain why scaling corrections appear to be well
described by the $\mathcal{O}(L^{-\omega})$ behavior practically for all the values of $L$ 
for $\sigma^2=0.1$ but not for $\sigma^2=1$: since the size of the
$\mathcal{O}(L^{-\omega})$ corrections for $\sigma^2=1$ is smaller than for
$\sigma^2=0.1$, the effect of higher dimensional irrelevant operators
becomes noticeable in the $\sigma^2=1$ case.

A completely analogous analysis has been carried out also for the single site
Metropolis update with the truncation error described by
Eq.~\eqref{eq:metro_rand}. Also in this case we considered three values of the
cut-off $\Lambda$ ($\Lambda=10^{3}, 10, 2$), and for each value of $\Lambda$ we
performed simulation at several values of $\beta$ using four different lattice
sizes ($L=8, 16, 24, 32$). 

The critical temperature has been estimated, for the different values of the
cut-off $\Lambda$ investigated, using the same procedure described above, and
the final results are reported in Tab.~\ref{tab:betac_metro_cut}.  Also in this
case the scaling of $R_{\xi}$ as a function of $(\beta-\beta_c)L^{1/\nu}$,
obtained by fixing the value of $\nu$ corresponding to the XY universality
class, is extremely good even for the huge value $\Lambda=2$ of the 
cut-off, as can be seen from Fig.~\ref{fig:metro_cut_lambda2_1}.

A further check of the XY nature of the critical behavior is provided by the
scaling of $U$ as a function of $R_{\xi}$: also in this case data converge to
the universal curve $\mathcal{U}(R_{\xi})$ of the XY model determined in
Sec.~\ref{sec:XYscal}, and in Fig.~\ref{fig:metro_cut_lambda2_2} we report the
results for $L^{\omega}(U(R_{\xi})-\mathcal{U}(R_{\xi}))$. The fact that
numerical results for this quantity asymptotically (in the $L\to\infty$ limit)
collapse on a scaling curve ensures that $U(R_{\xi})$ is effectively converging
to the universal XY scaling function $\mathcal{U}(R_{\xi})$ with
$O(L^{-\omega})$ corrections. By comparing the results reported in
Fig.~\ref{fig:metro_cut_lambda2_2} for $\Lambda=10$ and $\Lambda=2$ (note the
different scale on the vertical axis in the two panels) we can see that by
decreasing $\Lambda$, i.~e. by increasing the size of the truncation error, the
size of the leading corrections to scaling decreases, which is the equivalent
in the present case of what already noted when discussing the stochastic
perturbation of the single site Metropolis algorithm.

\subsubsection{Single cluster update}

\begin{table}[t]
\begin{tabular}{lll}
\hline\hline
$\sigma^2$ & $\beta_c(R)$   & $\beta_c(U)$ \\ \hline
0.1        & 0.4887867(44)  & 0.4887893(55) \\ \hline  
0.5        & 0.608378(17)   & 0.608361(24)  \\ \hline\hline
\end{tabular}
\caption{Single cluster update with Gaussian random energy perturbation of the
cluster formation probability (see Eq.~\eqref{eq:cluster_rand}): estimates of
the critical temperature for different values of $\sigma^2$, obtained by
analyzing $R_{\xi}$ or $U$. For comparison the critical temperature of the XY
model is 0.45416474(10)[7] (from Ref.~\cite{Hasenbusch-19}).
Note that the critical temperature is not universal, and there is
no reason to expect $\beta_c$ not to vary with $\sigma^2$. Its variation is in
fact an indication that the values of $\sigma^2$ used are large enough to have an effect on the system.
}
\label{tab:betac_cluster_rand}
\end{table}

\begin{table}[t]
\begin{tabular}{lll}
\hline\hline
$\Lambda$ & $\beta_c(R)$   & $\beta_c(U)$ \\ \hline
$10^3$    & 0.4545794(11)  & 0.4545782(10) \\ \hline
$10^2$    & 0.4583497(15)  & 0.4583490(15) \\ \hline
10        & 0.5004(2)      & 0.5007(2)     \\  \hline\hline
\end{tabular}
\caption{Single cluster update with truncation perturbation of the cluster
formation probability (see Eq.~\eqref{eq:cluster_cut}): estimates of the
critical temperature for different values of $\Lambda$, obtained by analyzing
$R_{\xi}$ or $U$. For comparison the critical temperature of the XY model is
0.45416474(10)[7] (from Ref.~\cite{Hasenbusch-19}).
Note that the critical temperature is not universal, and there is
no reason to expect $\beta_c$ not to vary with $\Lambda$. Its variation is in
fact an indication that the values of $\Lambda$ used are large enough to have an effect on the system.
}
\label{tab:betac_cluster_cut}
\end{table}

Let us now discuss how MCMC perturbations affect the result obtained by using
the single cluster algorithm.  We start again from the stochastic perturbation
case, and in the cluster algorithm the perturbation involves the acceptance
probability to be used in building the cluster. In particular we consider the
Gaussian perturbation in Eq.~\eqref{eq:cluster_rand} of the standard acceptance
probability Eq.~\eqref{eq:cluster}, with zero average and variance $\sigma^2$.
In this case we considered two values of the variance ($\sigma^2=0.1, 0.5$),
and for each value of $\sigma^2$ we performed simulations for several $\beta$
values, using lattices of linear extent $L=8,16,24,32$. 

\begin{figure}[t]
\includegraphics[width=0.9\columnwidth, clip]{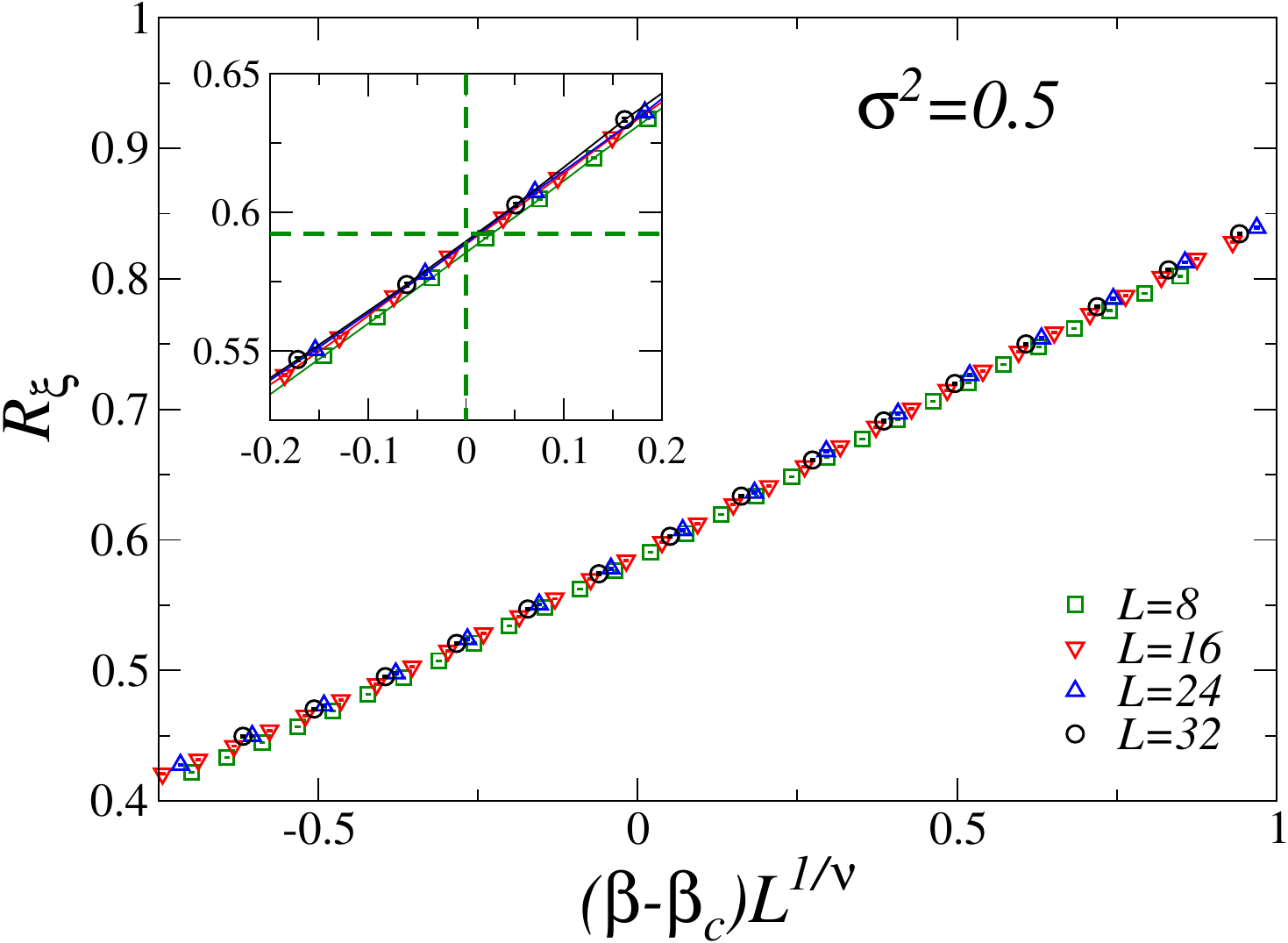}
\caption{Results for the single cluster update with Gaussian random
perturbation (see Eq.~\eqref{eq:cluster_rand}) with variance $\sigma^2=0.5$.
FSS of $R_{\xi}$ obtained by using the value of the critical temperature
reported in Tab.~\ref{tab:betac_cluster_rand} and the value of the critical
exponent $\nu$ reported in Tab.~\ref{tab:XY}.  The inset shows a zoom of the
region close to $\beta\approx \beta_c$, and dashed lines are drawn in
correspondence of $\beta=\beta_c$ and $R_{\xi}=R_{\xi}^*$. 
} 
\label{fig:cluster_rand_sigma0p5_1}
\end{figure}

\begin{figure}[t]
\includegraphics[width=0.9\columnwidth, clip]{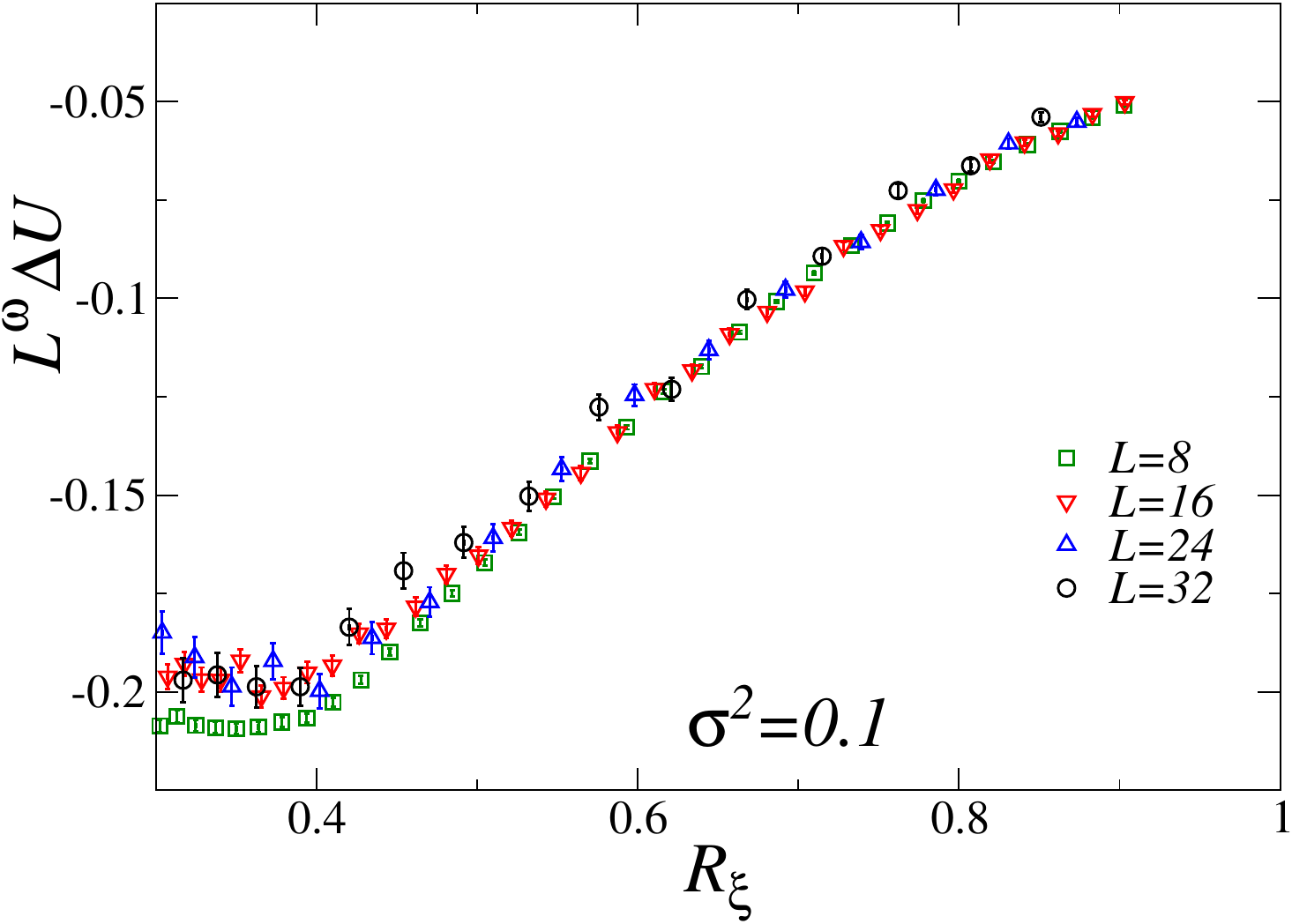}\\
\includegraphics[width=0.9\columnwidth, clip]{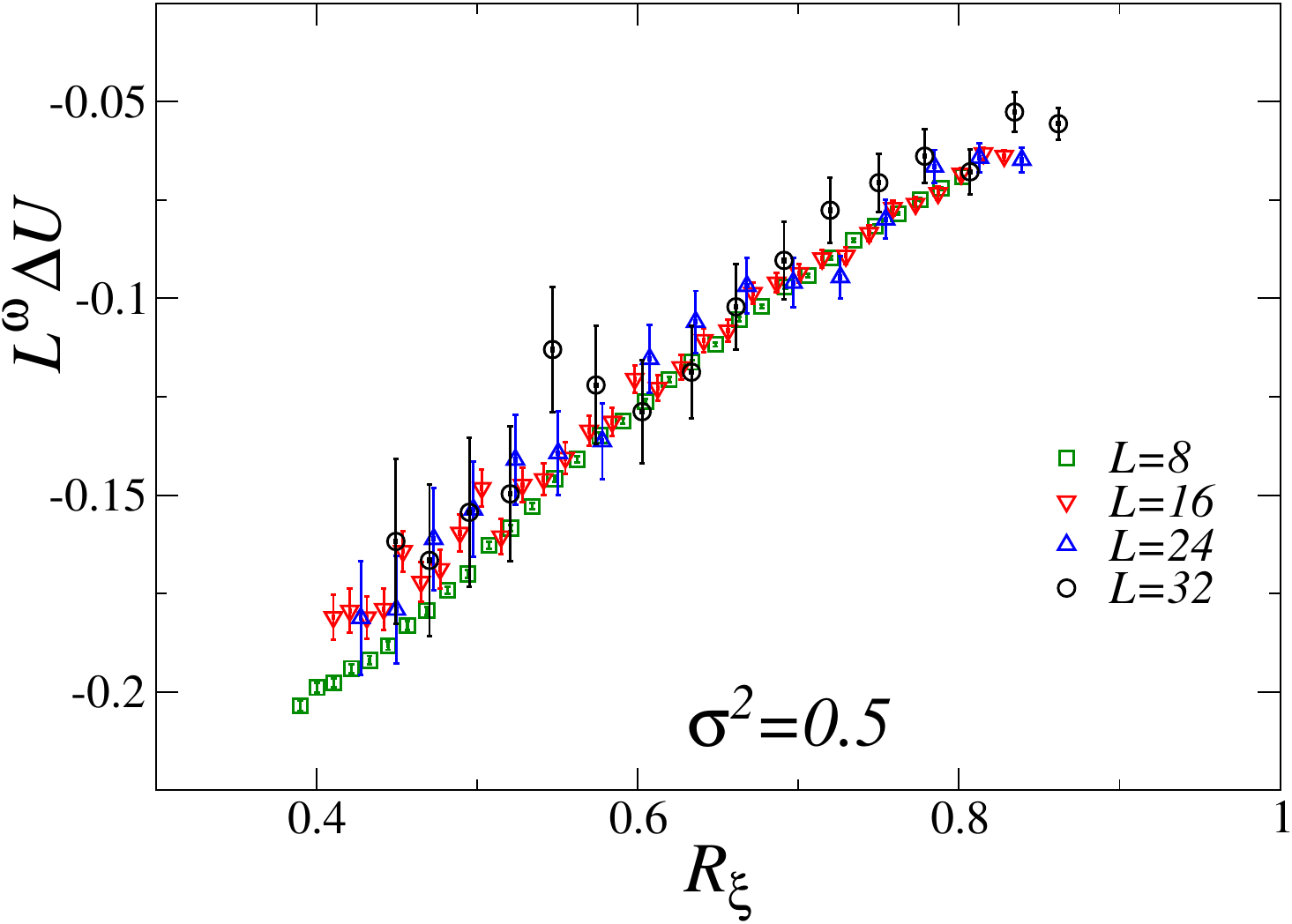}
\caption{Results for the single cluster update with Gaussian random
perturbation (see Eq.~\eqref{eq:cluster_rand}). Estimates of
$L^{\omega}(U(R_{\xi})-\mathcal{U}(R_{\xi}))$, for the model with
$\sigma^2=0.1$ (upper panel) and $\sigma^2=0.5$ (lower panel).}
\label{fig:cluster_rand_sigma0p5_2}
\end{figure}

Critical temperatures have been determined using the same procedure described
when discussing the single site Metropolis perturbations, and the final results
obtained in this way are reported in Tab.~\ref{tab:betac_cluster_rand}.
Deviations of the critical temperatures in Tab.~\ref{tab:betac_cluster_rand}
from the $\beta_c$ value of the unperturbed XY model are larger than for the
previously studied perturbations (see Tabs.~\ref{tab:betac_metro_rand},
\ref{tab:betac_metro_cut}), however the critical behavior turns out to be still 
consistent with that of the three dimensional XY universality class.  

In Fig.~\ref{fig:cluster_rand_sigma0p5_1} we report $R_{\xi}$ data as a
function of $(\beta-\beta_c)L^{1/\nu}$ for the case $\sigma^2=0.5$, where
$\beta_c$ can be read from Tab.~\ref{tab:betac_cluster_rand} and for $\nu$ we
used the XY universality class value in Tab.~\ref{tab:XY}.  Some slight scaling
corrections are present in data computed on the $L=8$ lattice, but overall the
data collapse is still very good on larger lattices, and a similar picture
emerges from the analysis of $U$ data as a function of
$(\beta-\beta_c)L^{1/\nu}$ (not shown).

Results for $L^{\omega}(U(R_{\xi})-\mathcal{U}(R_{\xi}))$ as a function of
$R_{\xi}$ are shown in Fig.~\ref{fig:cluster_rand_sigma0p5_2} for the values
$\sigma^2=0.1$ and $\sigma^2=0.5$ of the noise variance. These data clearly show that
the universal scaling function $\mathcal{U}(R_{\xi})$ correctly describes the
asymptotic FSS behavior of $U$ as a function of $R_{\xi}$, with
$O(L^{-\omega})$ corrections and $\omega=\omega_{XY}$, thus further confirming
that the observed critical behavior is consistent with that of the XY
universality class. For the cluster update with a stochastic perturbation also
the size of the scaling corrections is almost independent of the variance
$\sigma^2$, as can be seen by comparing data in
Fig.~\ref{fig:cluster_rand_sigma0p5_2} with those in Fig.~\ref{fig:fit_XY}
(lower panel). It is however important to stress that errors do change with the
value of $\sigma^2$: data reported in the two panels of
Fig.~\ref{fig:cluster_rand_sigma0p5_2} have been obtained using similar
statistics (differing at most by a factor of $2$) but the errors observed for
$\sigma^2=0.5$ are much larger than what could be expected by rescaling the
errors for $\sigma^2=0.1$. This is an
indication of the fact that the perturbation is not strong enough to break down
the algorithm (the correct universal critical behavior is still observed) but
the correlation between the clusters identified by the algorithm and the
relevant critical modes decreases as $\sigma^2$ is increased.

\begin{figure}[t]
\includegraphics[width=0.9\columnwidth, clip]{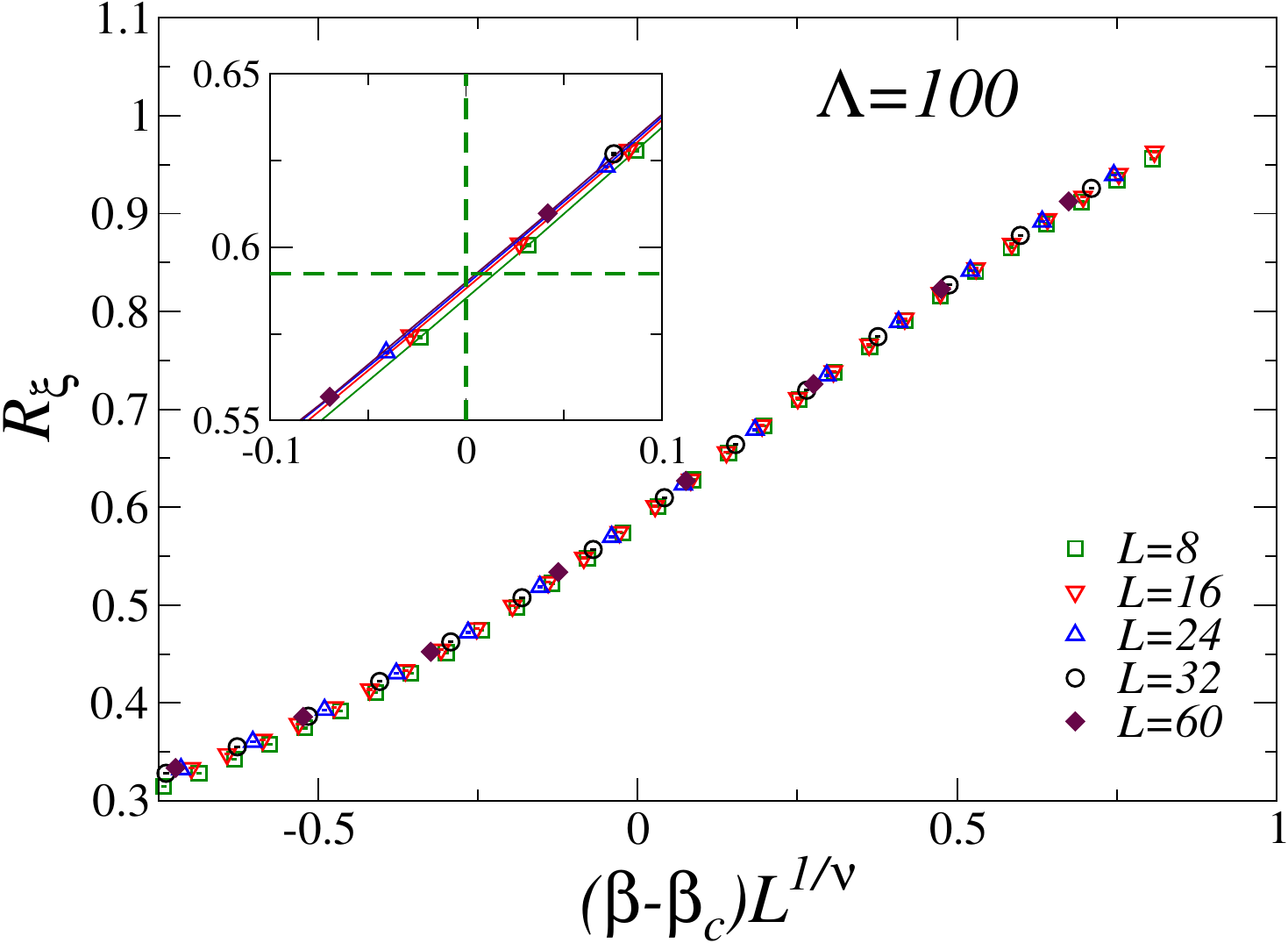}\\
\includegraphics[width=0.9\columnwidth, clip]{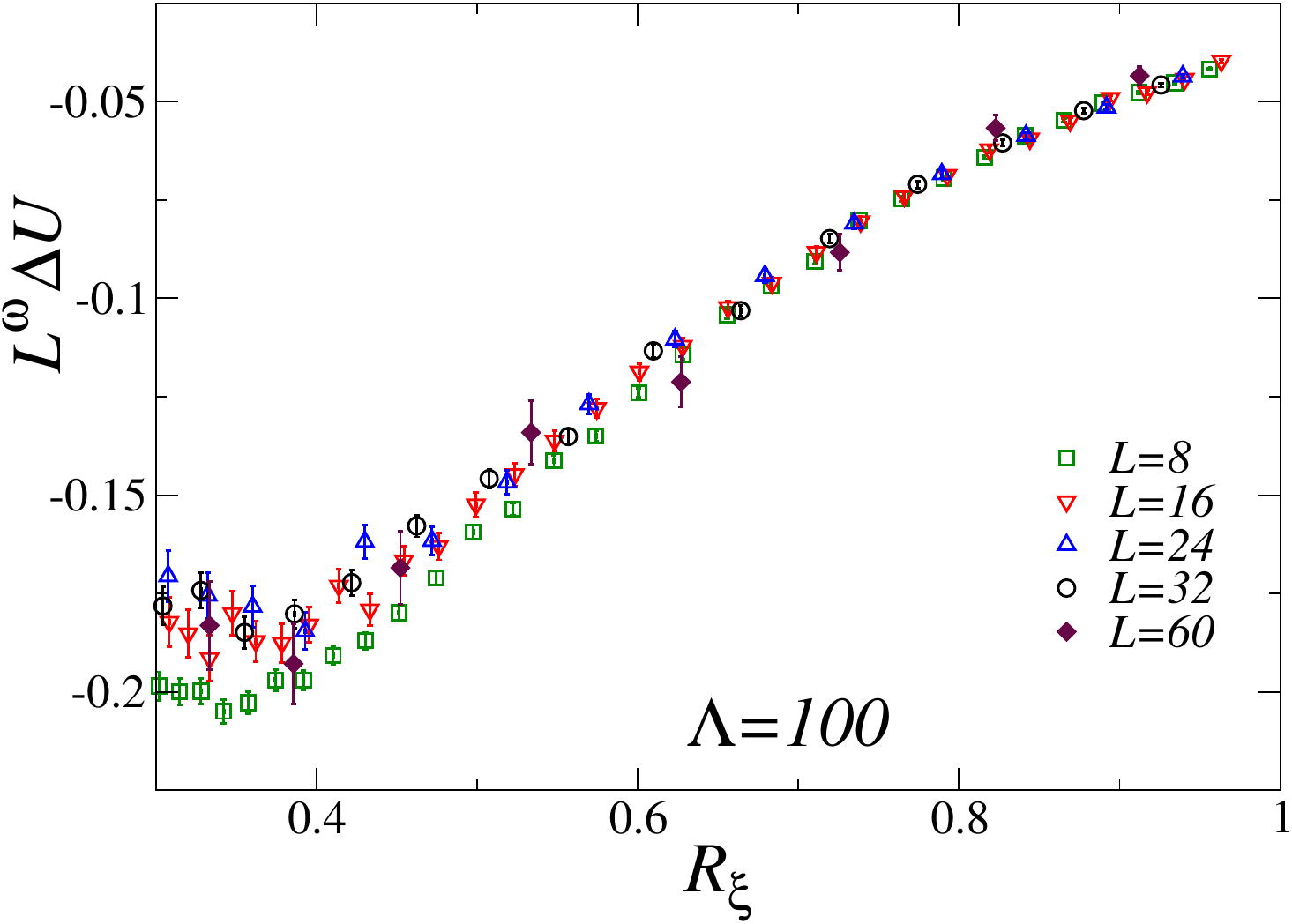}
\caption{Results for the single cluster update with truncation energy
perturbation (see Eq.~\eqref{eq:cluster_cut}) and $\Lambda=100$. (upper panel)
FSS of $R_{\xi}$ obtained by using the value of the critical temperature
reported in Tab.~\ref{tab:betac_cluster_cut} and the value of the critical
exponent $\nu$ reported in Tab.~\ref{tab:XY}.  The inset shows a zoom of the
region close to $\beta\approx \beta_c$, and dashed lines are drawn in
correspondence of $\beta=\beta_c$ and $R_{\xi}=R_{\xi}^*$.  (lower panel)
Estimates of $L^{\omega}(U(R_{\xi})-\mathcal{U}(R_{\xi}))$, with
$\omega=\omega_{XY}$, see Tab.~\ref{tab:XY}.
} 
\label{fig:cluster_cut_lambda100}
\end{figure}

\begin{figure}[t]
\includegraphics[width=0.9\columnwidth, clip]{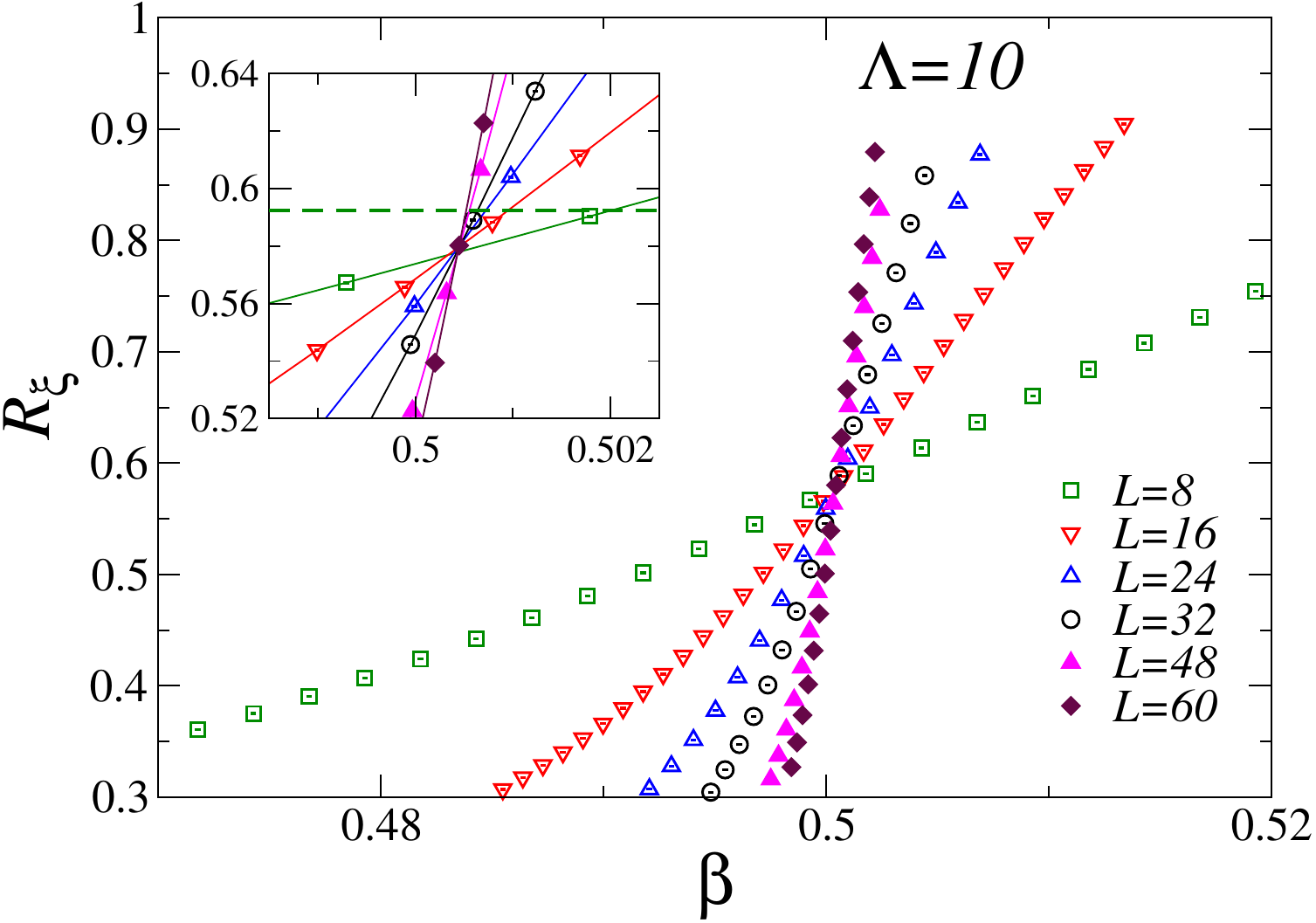}\\
\includegraphics[width=0.9\columnwidth, clip]{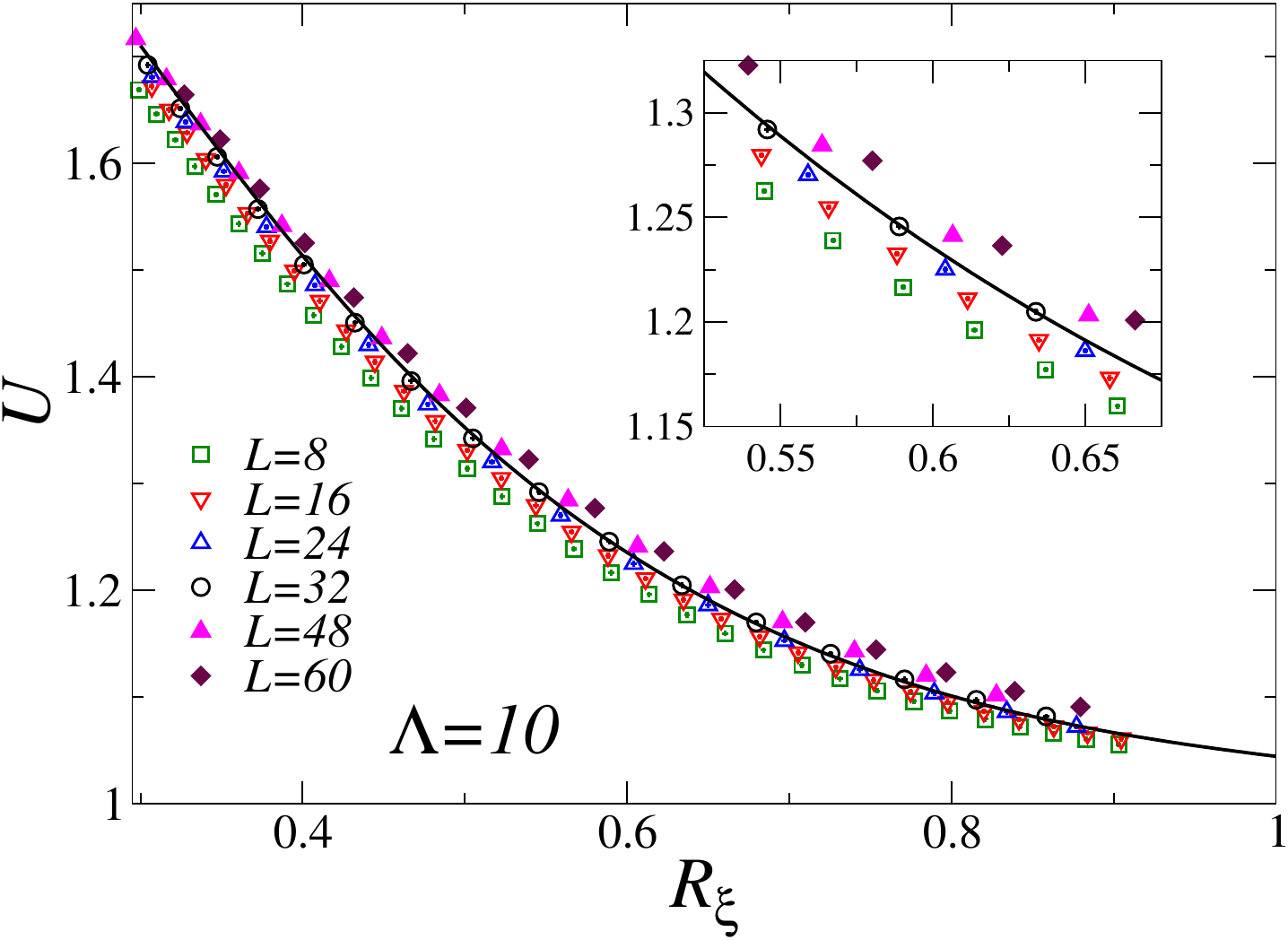}
\caption{Results for the single cluster update with truncation energy
perturbation (see Eq.~\eqref{eq:cluster_cut}) and $\Lambda=10$. (upper panel)
Data for $R_{\xi}$; in the inset a zoom of the $\beta\approx\beta_c$ region is
shown, and the horizontal dashed line is drawn in correspondence of the
universal XY critical value  $R_{\xi}^*$ reported in Tab.~\ref{tab:XY}.  (lower
panel) Data for $U$ against $R_{\xi}$, plotted together with the universal scaling
curve $\mathcal{U}(R_{\xi})$ of the XY model obtained in Sec.~\ref{sec:XYscal}.
}
\label{fig:cluster_cut_lambda10}
\end{figure}

Let us finally consider the case of the single cluster update with a truncation
error in the probability of adding a site to the cluster, see
Eq.~\eqref{eq:cluster_cut}. To study this case we considered three values of
the cut-off $\Lambda$ ($\Lambda=10^3, 10^2, 10$), and for each value of
$\Lambda$ we performed simulations for several $\beta$ values, using lattices of
linear extent going from $L=8$ up to $L=60$.

For $\Lambda=10^3$ and $\Lambda=10^2$ it was possible to carry out the analysis
as done in the previous cases: the critical temperatures have been
estimated by performing biased fits of $R_{\xi}$ and $U$ data, using 
polynomial approximations of the FSS relation Eq.~\eqref{eq:FSS_RG} with the
constants $\nu, \omega, R_{\xi}^*$ and $U^*$ fixed to the universal XY values
reported in Tab.~\ref{tab:XY}.  The final results for the critical temperatures
obtained in this way are reported in Tab.~\ref{tab:betac_cluster_cut}.  For
both the $\Lambda=10^3$ and $\Lambda=10^2$ cases the critical behavior is fully
consistent with the standard XY universality class. Results for $\Lambda=10^2$
are shown in Fig.~\ref{fig:cluster_cut_lambda100}, from which it is clear that
$R_{\xi}$ scales with $(\beta-\beta_c)L^{1/\nu}$, $\nu=\nu_{XY}$, and that the
asymptotic behavior of $U$ as a function of $R_{\xi}$ is consistent with the
scaling function $\mathcal{U}(R_{\xi})$ determined in  Sec.~\ref{sec:XYscal}.

The case $\Lambda=10$ is the only one in which biased fits to extract the
critical temperature where not able to correctly reproduce the behavior of
$R_{\xi}$ and $U$ data.  From Fig.~\ref{fig:cluster_cut_lambda10} (upper panel)
it is indeed quite clear that the crossing point of $R_{\xi}$ data is different
from the value $R_{\xi}^*$ reported in Tab.~\ref{tab:XY}.  Analogously, the XY
scaling curve $\mathcal{U}(R_{\xi})$ does not describe the asymptotic behavior
of $U$ as function of $R_{\xi}$, as can be clearly seen from the lower panel of
Fig.~\ref{fig:cluster_cut_lambda10}. For $\Lambda=10$ the transition is thus
definitely not in the three dimensional XY universality class.

The values of the critical temperature reported in
Tab.~\ref{tab:betac_cluster_cut} for the case $\Lambda=10$ have been obtained
by performing unbiased fits of $R_{\xi}$ and $U$, i.~e. using the values of
$\nu$, $R_{\xi}^*$ and $U^*$ as fit parameters, and considering several values
of $\omega$ in the range $[0,1]$. These fits did not provided us with a solid
estimate of the critical exponent $\nu$, which drift from $\approx 0.7$ to
$\approx 0.9$ when data obtained on small $L$ lattices are excluded from the
fit.  All these facts together contribute to make the estimates of $\beta_c$
in this case much less accurate than in the other cases.

\section{Conclusions}
\label{concl}

We investigated the stability of the universal properties at continuous
phase transitions against perturbations of the MCMC algorithm, considering both
local and global update schemes, and stochastic and deterministic
perturbations.  

For the case of the classical three dimensional XY model, which we used as test
bed for this investigation, we found universal properties to be surprisingly
stable, even against very large perturbations.  Indeed in all but one case
critical exponents and universal scaling curves have been unaffected, within 
the statistical accuracy, by the introduction of perturbations so large as to modify
the second significant figure of the energy.

It is natural to interpret such a stability as due to the presence of
``universality classes'' of MCMC algorithms, however the development of an
appropriate RG framework to handle these type of perturbation is obviously
required to put this interpretation on solid ground.  Note that the
framework required is different from that of critical dynamics
\cite{HH-77, FM-06}: for what concerns the stability against
perturbations local and global algorithms appear to behave in an analogous way,
although they definitely have different critical dynamical
properties (and in particular different values of the dynamic critical exponent $z$).
Despite all these precautions it is difficult not to state that the
perturbations considered in this work are irrelevant in a RG-like sense.

To further corroborate this interpretation of the numerical results more
investigations are clearly required, to check if a similar phenomenology is
observed also using different statistical models. It would in particular
be interesting to understand what happens at charged transitions in gauge theories
\cite{Bonati:2024sok}, since perturbations of the dynamics could result in soft
breaking of the gauge invariance.

Another possibility for further studies is the characterization of the critical
behavior emerging when the perturbation changes the universality class of the
transition.  We could identify only one case in which this happens, i.~e. the single
cluster update with a very large truncation error, but large scaling corrections
prevented us from characterizing the critical properties in this case. In
particular we have no hints if the critical behavior observed 
corresponds to an already known universality class or to something new.

For classical simulations performed on standard CPUs the topic investigated in
this work is probably important mainly from the theoretical point of view,
but it can be of direct practical relevance in the context of
quantum algorithms, and for simulations performed using low-accuracy hardware
or surrogate energy functions.

\acknowledgments

It is a pleasure to thank Francesco Grotto and Silvia Morlacchi for useful
discussions.  C.B. acknowledge support from the project PRIN 2022 ``Emerging
gauge theories: critical properties and quantum dynamics'' (20227JZKWP). This
study was carried out within the National Centre on HPC, Big Data and Quantum
Computing - SPOKE 10 (Quantum Computing) and received funding from the European
Union Next-GenerationEU - National Recovery and Resilience Plan (NRRP) --
MISSION 4 COMPONENT 2, INVESTMENT N. 1.4 -- CUP N.  I53C22000690001.  Numerical
simulations have been performed on the CSN4 cluster of the Scientific Computing
Center at INFN-PISA.

\end{document}